\documentclass{physeauth}
\usepackage{graphicx}
\usepackage{amsmath}
\usepackage{amssymb}
\usepackage{bm}

\begin{document}
\begin{frontmatter}

\title{Quantum decoherence of interacting electrons in arrays of quantum dots
  and diffusive conductors}
\author[int,fian]{Dmitri S. Golubev \thanksref{thank1}},
\author[int,fian]{Andrei D. Zaikin}

\address[int]{Forschungszentrum Karlsruhe, Institut f\"ur Nanotechnologie,
76021 Karlsruhe, Germany}
\address[fian]{I.E.Tamm Department of Theoretical Physics, P.N.Lebedev
Physics Institute, 119991 Moscow, Russia\\}
\thanks[thank1]{Corresponding author.  E-mail: golubev@int.fzk.de}

\begin{abstract}
We develop a new unified theoretical approach enabling us to
non-perturbatively study the effect of electron-electron
interactions on weak localization in arbitrary arrays of quantum
dots. Our model embraces (i) weakly disordered conductors
(ii) strongly disordered conductors and (iii) metallic quantum
dots. In all these cases at $T \to 0$ the electron decoherence
time is determined by the universal formula $\tau_{\varphi 0}\sim
g\tau_D/\ln (E_C/\delta)$, where $g$, $\tau_D$, $E_C$ and $\delta$
are respectively dimensionless conductance, dwell time, charging
energy and level spacing of a single dot. In the case (i) this
formula yields $\tau_{\varphi 0}\propto D^3/\ln D$ ($D$ is the
diffusion coefficient) and matches with our previous
quasiclassical results [D.S. Golubev, A.D. Zaikin, Phys. Rev.
Lett. 81 (1998) 1074], while in the cases (ii) and (iii) it
illustrates new physics not explored earlier. A detailed
comparison between our theory and numerous experiments provides an
overwhelming evidence that zero temperature electron decoherence in
disordered conductors is universally caused by electron-electron
interactions rather than by magnetic impurities.
\end{abstract}

\begin{keyword}
weak localization \sep decoherence \sep electron-electron
interactions \sep disorder \sep quantum dots
\PACS 73.63.Kv \sep
73.21.La \sep 73.20.Fz \sep 73.23.-b
\end{keyword}
\end{frontmatter}

\section{Introduction}

Quantum interference of electrons in mesoscopic conductors
manifests itself in a number of fundamentally important phenomena
which can be directly observed in modern experiments. One of them
is the phenomenon of weak localization (WL) \cite{Berg,AA,CS}. In
the absence of interactions electron wave functions preserve their
coherence and, hence, quantum interference remains efficient
throughout a large part of the sample making WL a pronounced
effect. Interactions between electrons and with other degrees of
freedom may limit phase coherence thereby making quantum
interference of electrons possible only within a finite length
scale $L_\varphi$. This so-called electron decoherence length as
well as directly related to it decoherence time $\tau_\varphi =
L_\varphi^2/D$ (where $D$ is diffusion coefficient) are crucial
parameters indicating importance of quantum effects in the system
under consideration.

At sufficiently high temperatures quantum behavior of electrons in
disordered conductors is usually suppressed due to various types
of interactions. However, as temperature gets lower, certain
interaction mechanisms either ``freeze out'' or become less
efficient in destroying quantum coherence. As a result, both
$L_\varphi$ and $\tau_\varphi$ usually grow with decreasing
temperature and quantum effects become progressively more
important.

Should one expect  $L_\varphi$ and $\tau_\varphi$ to diverge in
the limit $T \to 0$? While some authors tend to give a positive
answer to this question, numerous experiments performed on
virtually all kinds of disordered conductors and in all dimensions
demonstrate just the opposite, i.e. that at low enough $T$ both
decoherence length and time {\it saturate} to a constant and do
not anymore grow if temperature decreases further. The list of
corresponding structures and experiments, by far incomplete,
includes quasi-1d metallic wires
\cite{Moh,MJW,Nat,Saclay,MW,Ba,Bird,Nat05,Gre,Jap,Gre-new,Birge,many},
quasi-1d semiconductors \cite{Pooke,Nog,Kha}, carbon nanotubes
\cite{Leuwen,Kang,LTL}, 2d metallic
\cite{MJW,Nat,LG,Sah,Enh,BL,Lin07} and semiconductor
\cite{Pud,Sav,Sivan} films, various 3d disordered metals
\cite{BL,Lin07,Lin01} and (0d) quantum dots
\cite{Bird1,Clarke,Pivin,Huibers,Hackens}. Though dimensions and
parameters of these systems are different, the low temperature
saturation of $\tau_\varphi$ remains the common feature of all
these observations.

Is this ubiquitous saturation of $\tau_\varphi$ an {\it intrinsic}
or {\it extrinsic} effect? If intrinsic, decoherence of electrons
at $T=0$ would be a fundamentally important conclusion which would
shed a new light on the physical nature of the ground state of
disordered conductors as well as on their low temperature
transport properties. While extrinsic saturation of $\tau_\varphi$
could be caused by a variety of reasons, the choice of intrinsic
dephasing mechanisms is, in fact, much more restricted. There
exists, however, at least one mechanism, electron-electron
interactions, which remains important down to lowest temperatures
and may destroy quantum interference of electrons even at $T=0$
\cite{Moh,GZ98}.

In a series of papers \cite{GZ98,GZ99,GZ00,GZ03} we offered a
theoretical approach that allows to describe electron interference
effects in the presence of disorder and electron-electron
interactions at any temperature including the most interesting
limit $T \to 0$. This formalism extends Chakravarty-Schmid
description \cite{CS} of WL and generalizes
Feynman-Vernon-Caldeira-Leggett path integral influence functional
technique \cite{FH,CL,SZ,W} to fermionic systems with disorder and
interactions. With the aid of our approach we have evaluated WL
correction to conductance and electron decoherence time in the
limit $T \to 0$ and demonstrated that low temperature saturation
of $\tau_\varphi$ can indeed be caused by electron-electron
interactions. Our results allowed for a direct comparison with
experiments and a good agreement between our theory and numerous
experimental data for $\tau_\varphi$ in the low temperature limit
was found \cite{GZ98,GZ99,Lammi,GZ02}. In particular, for quasi-1d
wires with thicknesses exceeding the elastic electron mean free
path $l$ at $T \to 0$ our theory predicts $\tau_\varphi \propto
D^3$, where $D$ is the diffusion coefficient. This scaling is
indeed observed in experiments for not very strongly disordered
wires typically with $D \gtrsim 10$ cm$^2$/s (see Sec. 6 for more
details).

On the other hand, for strongly disordered structures with smaller
values of $D$ this scaling is not anymore fulfilled and, moreover,
an opposite trend is observed: $\tau_\varphi$ was found {\it to
increase} with decreasing $D$ \cite{Lin07,Lin01,Lin07b}. This
trend is {\it not} described by our expressions for $\tau_\varphi$
\cite{GZ98,GZ99}. Another interesting scaling was observed in
quantum dots: saturated values $\tau_\varphi$ were argued
\cite{Hackens} to scale with the dot dwell times $\tau_D$ as
$\tau_\varphi \approx \tau_D$. Our theory \cite{GZ98,GZ99} cannot
be directly used in order to explain the latter scaling either.

In order to attempt to reconcile all these observations within one
approach it is necessary to develop a unified theoretical
description which would cover essentially all types of disordered
conductors. It is worth pointing out that the technique
\cite{GZ99,GZ00,GZ03} is formally an exact procedure which should
cover all situations. However, for some structures, such as, e.g.,
quantum dots and granular metals, it can be rather difficult to
directly evaluate the WL correction within this technique for the
following reasons.

First of all, our description in terms of quasiclassical electron
trajectories may become insufficient in the above cases, and
electron scattering on disorder should be treated on more general
footing. Another -- purely technical -- point is averaging over
disorder. In our approach \cite{GZ98,GZ99,GZ00,GZ03} it is
convenient to postpone disorder averaging until the last stage of
the calculation. In some cases -- like ones studied below -- it
might be, on the contrary, more appropriate to perform disorder
averaging already in the beginning of the whole consideration. In
addition, it is desirable to deal with the model which would
embrace various types of conductors with well defined properties
both in the long and short wavelength limits. This feature will
help to construct a fully self-contained theory free of any
divergencies and additional cutoff parameters.

Recently \cite{GZ06} we made a first step towards this unified
theory. Namely, we adopted a model for a disordered conductor
consisting of an array of (metallic) quantum dots connected via
junctions (scatterers) with arbitrary transmission distribution of
their conducting channels. This model allows to easily crossover
between the limits of a granular metal and that with point-like
impurities and to treat spatially restricted and spatially
extended conductors within the same theoretical framework, as
desired. Within this model in Ref. \cite{GZ06} we analyzed WL
corrections to conductance merely for non-interacting electrons
and included interaction effects by introducing the electron
dephasing time $\tau_\varphi$ just as a phenomenological
parameter. Systematic analysis of the effect of electron-electron
interactions on weak localization within this formalism will be
developed in this paper. This approach will allow to
microscopically evaluate $\tau_\varphi$ for all the structures
under consideration.

The structure of our paper is as follows. In Sec. 2 we will
discuss qualitative arguments illustrating the role of scattering
and interactions in electron dephasing. In Sec. 3 we introduce our
model of an array of quantum dots and outline a general
theoretical framework which is then employed in Sec. 4 and 5 for
rigorous calculations of the WL correction to conductance and
electron decoherence time in the presence of electron-electron
interactions. A detailed comparison of our results with numerous
experiments performed in various disordered conductors is carried
out in Sec. 6. A brief summary of our main results and conclusions
is presented in Sec. 7.

\section{Qualitative arguments}

Before turning to a detailed calculation it is instructive to
discuss a simple qualitative picture demonstrating under which
conditions electron dephasing by interaction is expected to occur.

\begin{figure}
\includegraphics[width=5.3cm]{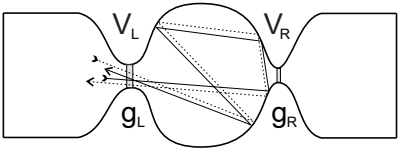}
\caption{Single quantum dot and a pair of time-reversed electron
paths. Fluctuating voltages $V_L$ and $V_R$ are assumed to drop
only across the barriers and not inside the dot.}
\end{figure}

Consider first the simplest system of two scatterers separated by
a cavity (quantum dot, Fig. 1) The WL correction to conductance
of a disordered system $G_{WL}$ is known to arise from
interference of pairs of time-reversed electron paths \cite{CS}.
In the absence of interactions for a single quantum dot of Fig. 1
this correction is evaluated in a general form \cite{B}. The
effect of electron-electron interactions can be described in terms
of fluctuating voltages. Let us assume that the voltage can drop
only across the barriers and consider two time-reversed electron
paths which cross the left barrier (with fluctuating voltage
$V_L(t)$) twice at times $t_i$ and $t_f$ as shown in Fig. 1. It
is easy to see that the voltage-dependent random phase factor
$\exp (i\int_{t_i}^{t_f}V_L(t)dt)$ acquired by the electron wave
function $\Psi$ along any path turns out to be exactly the same as
that for its time-reversed counterpart. Hence, in the product
$\Psi\Psi^*$ these random phases cancel each other and quantum
coherence of electrons remains fully preserved. This implies that
for the system of Fig. 1a fluctuating voltages (which can mediate
electron-electron interactions) {\it do not cause any dephasing}.

This qualitative conclusion can be verified by means of more
rigorous considerations. For instance, it was demonstrated
\cite{GZ041} that the scattering matrix of the system remains
unitary in the presence of electron-electron interactions, which
implies that the only effect of such interactions is transmission
renormalization but not electron decoherence. In Ref.
\cite{Brouwer} a similar conclusion was reached by directly
evaluating the WL correction to the system conductance. Thus, for
the system of two scatterers of Fig. 1a electron-electron
interactions can only yield energy dependent (logarithmic at
sufficiently low energies) renormalization of the dot channel
transmissions \cite{GZ01,BN} but not electron dephasing.

\begin{figure}
\includegraphics[width=7.5cm]{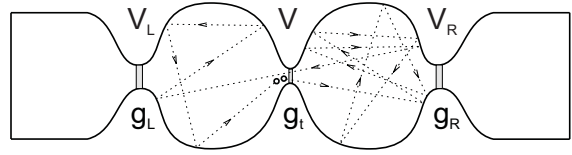}
\caption{Two quantum dots and a typical electron path. Fluctuating
voltages $V_L$, $V$ and $V_R$ are again assumed to drop only
across the barriers.}
\end{figure}

Let us now add one more scatterer and consider the system of two
quantum dots depicted in Fig. 2. We again assume that fluctuating
voltages are concentrated at the barriers and not inside the
cavities. The phase factor accumulated along the path (see Fig. 2)
which crosses the central barrier twice (at times $t_i$ and
$t>t_i$) and returns to the initial point (at a time $t_f$) is
$e^{i[\varphi (t_i)-\varphi (t)]}$, where $\dot\varphi /e=V(t)$ is
the fluctuating voltage across the central barrier. Similarly, the
phase factor picked up along the time-reversed path reads
$e^{i[\varphi (t_f+t_i-t)-\varphi (t_f)]}$. Hence, the overall
phase factor acquired by the product $\Psi\Psi^*$ for a pair of
time-reversed paths is $\exp (i \Phi_{\rm tot})$, where
$$
\Phi_{\rm tot}(t_i,t_f,t)=\varphi (t_i)-\varphi
(t)-\varphi^+(t_f+t_i-t)+\varphi (t_f).
$$
Averaging over phase fluctuations, which for simplicity are
assumed Gaussian, we obtain
\begin{eqnarray}
&&\left\langle e^{i \Phi_{tot}(t_i,t_f,t) } \right\rangle
=\,e^{-\frac{1}{2} \left\langle \Phi_{tot}^2(t_i,t_f,t)
\right\rangle} \nonumber\\ &&
=\,e^{-2F(t-t_i)-2F(t_f-t)+F(t_f-t_i)+F(t_f+t_i-2t)},
\label{phase}
\end{eqnarray}
where we defined the phase correlation function
\begin{equation}
F(t)=\langle (\varphi(t)-\varphi(0))^2\rangle /2. \label{F}
\end{equation}
Should this function grow with time the electron phase coherence
decays and, hence, $G_{WL}$ has to be suppressed below its
non-interacting value due to interaction-induced electron
decoherence.

The above arguments are, of course, not specific to systems with
three barriers only. They can also be applied to any system with
larger number of scatterers, i.e. virtually to any disordered
conductor where -- exactly for the same reasons -- one also
expects non-vanishing interaction-induced electron decoherence at
any temperature including $T=0$. In the next sections we will
develop a quantitative theory which will confirm and extend our
qualitative physical picture. We are going to give a complete
quantum mechanical analysis of the problem which fully accounts
for Fermi statistics of electrons and treats electron-electron
interactions in terms quantum fields produced internally by
fluctuating electrons. Below we will non-perturbatively evaluate
WL correction $G_{WL}$ for arrays of metallic quantum dots in the
presence of electron-electron interactions which will be shown to
reduce phase coherence of electrons at any temperature down to
$T=0$.

\section{The model and basic formalism}

\begin{figure}
\includegraphics[width=7.5cm]{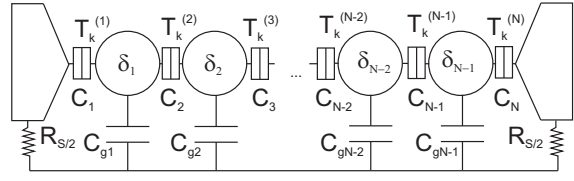}
\caption{Schematics of 1d array of $N-1$ quantum dots connected via
$N$ barriers. }
\end{figure}

Let us consider a 1d array of  $N-1$ quantum dots connected in
series chaotic quantum dots (Fig. 3). Each quantum dot is
characterized by its own mean level spacing $\delta_n$. Adjacent
quantum dots are connected to each other via barriers which can
scatter electrons. The first and the last dot are also connected
to the leads (with total resistance $R_S$), i.e. altogether we have $N$ scatterers in our
system. Each such scatterer is described by a set of transmissions
of its conducting channels $\tilde T_k^{(n)}$ (here $k$ labels the
channels and $n$ labels the scatterers).  Below we will focus our
attention on the case of metallic quantum dots with the level
spacing $\delta_n$ being the lowest energy parameter in the
problem.

The system of Fig. 3 will be described by the Hamiltonian
\begin{eqnarray}
\hat H&=&\sum_{n,m=1}^N\frac{C_{nm}\hat V_n\hat V_m}{2} +\hat
H_L+\hat H_R \nonumber\\ && +\sum_{n=1}^{N-1}\hat H_{n}^{\rm qd}
+\sum_{n=1}^N \hat H_n^{\rm T}, \label{H}
\end{eqnarray}
where $C_{nm}$ is the capacitance matrix of the array, $\hat V_n$
is the operator of electric potential on $n-$th quantum dot,
$$\hat
H_{L,R}=\sum_{k,\alpha=\uparrow,\downarrow} \hat
a^{L,R\;\dagger}_{\alpha,k} (\xi_k-eV_{L,R})\hat
a^{L,R}_{\alpha,k} $$
are the Hamiltonians of the left and right
leads, $V_{L,R}$ are the electric potentials of the leads fixed by
the external voltage source,
$$\hat H_n^{\rm
qd}=\sum_{kl,\alpha=\uparrow,\downarrow}\hat
a^{\dagger}_{n,\alpha,k}(H_{n,kl}^{\rm qd}-e\hat V_n)\hat
a_{n,\alpha,l}$$
is the Hamiltonian of $n$-th quantum dot and
$H_{n,kl}$ is a random matrix which belongs to the orthogonal
ensemble. Finally, electron transfer between adjacent $n-1$-th and $n$-th
quantum dots will
be described by the Hamiltonian $$\hat H_n^{\rm
T}=\sum_{\alpha=\uparrow,\downarrow}\int_J d^2{\bm r}\,
\big[t_n({\bm r})\hat\Psi^\dagger_{\alpha,n-1}({\bm r})
\hat\Psi_{\alpha,n}({\bm r})+{\rm c.c.}\big].$$ Here
the integration runs over the junction area.

Note that in Ref. \cite{GZ06} we have already applied the model of
Fig. 3 in order to analyze WL effects in the absence of
electron-electron interactions. In that paper we have used the
scattering matrix formalism combined with the non-linear
$\sigma$-model. In order to incorporate interaction effects into
our consideration it will be convenient for us to describe
inter-dot electron transfer within the tunneling Hamiltonian
approach, as specified above. We would like to emphasize that this
choice is only a matter of technical convenience, since the
approaches based on the tunneling Hamiltonian and on the
scattering matrix are fully equivalent to each other.

For the sake of completeness, let us briefly remind the reader the
relation between these two approaches. Consider, e.g., the $n$-th
barrier between the $n-1$-th and $n$-th quantum dots and define
the matrix elements $t_{lm}=\langle l|\hat H^{\rm T}_n| m\rangle$
between the $l-$th wave function in the $n-1$-th dot and  $m-$th
wave function in the $n$-th dot. Electron transfer across this
barrier can then be described by the set of eigenvalues of this
matrix $\tilde t_k$. These eigenvalues are related to the barrier
channel transmissions $\tilde T_k^{(n)}$ as ( see, e.g., \cite{B})
\begin{equation}
\tilde T_k^{(n)}=\frac{4\pi^2|\tilde t_k|^2/\delta_{n-1}\delta_n}
{(1+\pi^2|\tilde t_k|^2/\delta_{n-1}\delta_n)^2}.
\end{equation}
This equation allows to keep track of the relation between two
approaches at every stage of the calculation.

To proceed we will make use of the path integral Keldysh
technique. A brief sketch of this approach is outlined below. The
time evolution of the density matrix of our system is described by
the standard equation
\begin{equation}
\hat \rho(t)=e^{-i\hat Ht}\hat\rho_0\,e^{i\hat Ht},
\end{equation}
where $\hat H$ is given by Eq. (\ref{H}). Let us express the
operators $e^{-i\hat Ht}$ and $e^{i\hat Ht}$ via path integrals
over the fluctuating electric potentials $V_n^{F,B}$ defined
respectively on the forward and backward parts of the Keldysh
contour:
\begin{eqnarray}
e^{-i\hat Ht}&=&\int  DV_n^F\; {\rm T}\,\exp\left\{-i\int_0^t
dt'\hat H\left[V_n^F(t')\right]\right\},
\nonumber\\
e^{i\hat Ht}&=&\int  DV_n^B\; \tilde{\rm T}\,\exp\left\{i\int_0^t
dt'\hat H\left[V_n^B(t')\right]\right\}.
\end{eqnarray}
Here ${\rm T}\,\exp$ ($\tilde {\rm T}\,\exp$) stands for the
time ordered (anti-ordered) exponent and the
Hamiltonians $\hat H\left[V_n^F(t')\right]$, $\hat
H\left[V_n^B(t')\right]$ are obtained from the original Hamiltonian (\ref{H})
if one replaces the operators $\hat V_n(t)$ respectively
by the fluctuating voltages $V^F_n(t')$ and $V^B_n(t')$.

The central object of our analysis is the effective action
defined as
\begin{eqnarray}
iS[V^F,V^B]&=&\ln\left( {\rm tr} \left[ {\rm
T}\,\exp\left\{-i\int_0^t dt'\hat H\left[V_n^F(t')\right]\right\}
\right.\right. \nonumber\\ &&\times\, \left.\left. \hat\rho_0
\tilde{\rm T}\,\exp\left\{i\int_0^t dt'\hat
H\left[V_n^B(t')\right]\right\} \right]\right).
\end{eqnarray}
Since the operators $\hat H\left[V_n^F(t')\right]$, $\hat
H\left[V_n^B(t')\right]$ are quadratic in the electron creation and
annihilation operators, it is possible to rewrite the action in the form
\begin{equation}
iS=2\,{\rm tr}\,\ln\left[\check G^{-1}\right], \label{S}
\end{equation}
where $\check G^{-1}$ is the inverse Keldysh Green function of the system. The
operator  $\check G^{-1}$ has the following matrix structure
\begin{eqnarray}
\check G^{-1}=\left(
\begin{array}{ccccc}
\check G_L^{-1} & \check T_1 & 0 & 0 &\dots \\
\check T_n^\dagger & \check G_1^{-1} & \check T_2 & 0 & \dots\\
0 & \check T_2^\dagger & \check G_2^{-1} & \check T_3 \\
&& \ddots & & \\
\dots & 0 & \check T_{N-1}^\dagger & \check G_{N-1}^{-1} & \check T_N \\
\dots & 0 & 0 & \check T_N^\dagger & \check G_R^{-1}
\end{array}
\right). \label{invG}
\end{eqnarray}
Here each quantum dot as well as each of the two leads is represented by a
$2\times 2$ diagonal block
\begin{eqnarray}
\check G_n^{-1}=\left(
\begin{array}{cc}
i\partial_t -H^{\rm qd}_n+eV_n^F & 0 \\
0 & -i\partial_t +H^{\rm qd}_n-eV_n^B
\end{array}
\right),
\end{eqnarray}
while barriers are described by off-diagonal blocks $\check T_n$,
which have the form
\begin{equation}
\check T_n=\left(
\begin{array}{cc}
-t_n(\bm{r}) & 0 \\
0 & t_n(\bm{r})
\end{array}
\right).
\end{equation}
Below we will employ these general expressions in specific
situations of two quantum dots and of an array with a large number
of quantum dots. The latter system will also serve as a model for
spatially extended diffusive conductors.

\section{System of three barriers}

Let us first consider the system of three scatterers $N=3$
(two quantum dots). This system is important both in the context of electron
dephasing in quantum dots \cite{Bird1,Clarke,Pivin,Huibers,Hackens}
and as a relatively simple example which
illustrates all significant features of the effect of
electron-electron interactions on WL in more complicated systems.

Here we will treat interaction effects practically {\it without
any approximations}. For the sake of simplicity we will assume
that the central (i.e. second) barrier is a tunnel junction with
$T_k\ll 1$ for all its transmission channels. This assumption
allows one to expand the action (\ref{S}) in powers of the
parameter $t_2(\bm{r})$. We will also assume that dimensionless
conductance of the central barrier $g_2=2\pi/e^2R_2$ is much lower
than those of the first and third barriers which, in turn,
strongly exceed unity, i.e, $g_1,g_3 \gg g_2, 1$.  The effective
expansion parameter in this case is $g_2^2/g_1g_3\ll 1$. Then the
WL correction to the conductance in the absence of interactions
takes a particularly simple form \cite{GZ06}. E.g., for fully open
outer barriers we have
\begin{eqnarray}
G_{WL}^{\rm array}&=&
-\frac{e^2 g_2^2}{\pi}\left(\frac{1}{g_1^2}+\frac{1}{g_3^2}\right)+G_{WL}^{(0)},
\label{G}
\\
G_{WL}^{(0)}&=&-\frac{2e^2}{\pi}\frac{g_2^2}{g_1g_3}.
\label{G0}
\end{eqnarray}
The first term in Eq. (\ref{G}) defines the contribution of the
first and third scatterers, while the term $G_{WL}^{(0)}$ comes
from the central junction. Experimentally one can access
$G_{WL}^{(0)}$ attaching additional voltage probes to the quantum
dots. The aim of our subsequent analysis is
to demonstrate how the result (\ref{G0}) is modified in the
presence of electron-electron interactions.

\subsection{WL correction in the presence of interactions}

Setting $N=3$ in Eq. (\ref{H}) and defining the fluctuating phases
across the barriers $\varphi_n^{F,B}=\int_{t_0}^t
dt'\,\big[eV_n^{F,B}(t')- eV_{n-1}^{F,B}(t')\big]$ we perform the
gauge transformation which yields the new expressions for the
blocks $\check G_n^{-1}$ and $\check T_n$:
\begin{equation}
\check G_n^{-1}\to \left(
\begin{array}{cc}
i\partial_t -H^{\rm qd}_n & 0 \\
0 & -i\partial_t +H^{\rm qd}_n
\end{array}
\right),
\end{equation}
\begin{equation}
\check T_n\to \left(
\begin{array}{cc}
-t_n(\bm{r})e^{-i\varphi_n^F} & 0 \\
0 & t_n(\bm{r})e^{-i\varphi_n^B}
\end{array}
\right).
\end{equation}
Since the central barrier transmission is small we can
expand the action (\ref{S}) in powers of the parameter
$t_2(\bm r)$. Proceeding with this expansion up to the fourth order we get
\begin{equation}
iS=iS_S[\varphi_{1,3}^{F,B}]+iS_2^{(2)}+iS_2^{(4)}.
\end{equation}
Here
\begin{equation}
iS_S[\varphi_{1,3}^{F,B}]=iS_C+iS_{\rm ext} +2\,{\rm
tr}\,\ln[\check G^{-1}]\big|_{t_2(\bm{r})=0}
\label{Ss}
\end{equation}
is the effective action at zero
transmission of the second barrier which also includes terms
describing capacitances ($iS_C$) and the external circuit
($iS_{\rm ext}$), the term $iS_2^{(2)}\propto t_2^2(\bm r)$ is the
standard Ambegaokar-Eckern-Sch\"on (AES) action \cite{SZ} and
$iS_2^{(4)}\propto t_2^4(\bm r)$ contains information which will
allow us to evaluate the WL correction to the system conductance.
The corresponding expression reads
\begin{eqnarray}
 iS^{(4)}_2&=& -\sum_{i,j,k,l=F,B}\int dt_1\dots dt_4 \int_J d\bm{x}_1\dots d\bm{x}_4\,
\nonumber\\ &&\times\, \check G_{1,ij}(X_1; X_2)(-1)^j
e^{-i\varphi^j_2(t_2)}t(\bm{x}_2) \nonumber\\ && \times\, \check
G_{2,jk}(X_2;X_3)(-1)^k e^{i\varphi^k_2(t_3)}t(\bm{x}_3) \nonumber\\
&&\times\, \check G_{1,kl}(X_3; X_4)(-1)^l
e^{-i\varphi^l_2(t_4)}t(\bm{x}_4) \nonumber\\ &&\times\, \check
G_{2,li}(X_4;X_1)(-1)^i e^{i\varphi^i_2(t_1)}t(\bm{x}_1). \label{S2}
\end{eqnarray}
Here we use the convention $(-1)^{F}=-1$, $(-1)^B=1$,
$X=(t,\bm{x})$, and $\check G_{r}$ are $2\times 2$ matrix
Green-Keldysh functions in the first and the second quantum dots
($r=1,2$):
\begin{eqnarray}
i\check G_r=\left(
\begin{array}{cc}
\langle {\rm T}\,\hat\psi_r(X_1)\hat\psi^\dagger_r(X_2) \rangle &
-\langle \hat\psi^\dagger_r(X_2)\hat\psi_r(X_1) \rangle \\
 \langle \hat\psi_r(X_1)\hat\psi^\dagger_r(X_2) \rangle &
\langle \tilde{\rm T}\,\hat\psi_r(X_1)\hat\psi^\dagger_r(X_2) \rangle
\end{array}
\right).
\end{eqnarray}
Here we will set the Green functions $\check G_1$ and $\check G_2$
equal to their equilibrium values $\check G_r=G_r^R\check
F_1-\check F_2G_r^A$, where $G^{R,A}_r$ are retarded and advanced
Green functions,
\begin{equation}
\check F_1= \left(
\begin{matrix} h(E)& -f(E)\\ h(E) &
-f(E)\end{matrix}\right), \;\;\;\check F_2= \left(
\begin{matrix} f(E)& f(E)\\ -h(E) &
-h(E)\end{matrix}\right), \label{distr}
\end{equation}
$f(E)$ is the Fermi function and $h(E)=1-f(E)$. This choice is
sufficiently accurate for the problem in question. We will return
to this point towards the end of this section.

Our next step amounts to averaging the products of retarded and
advanced propagators in the action (\ref{S2}) over disorder in
each of the two dots separately. This averaging can be
accomplished, e.g., with the aid of the non-linear $\sigma$-model
or by other means. For the first (left) dot we obtain (cf., e.g.,
\cite{Tan})
\begin{eqnarray}
&&\langle  G^R_1(X_1,X_2)G^A_1(X_3,X_4) \rangle =
\nonumber\\
&& 2\pi
N_1{\mathcal V}_1 w(|\bm{r}_1-\bm{r}_4|)w(|\bm{r}_2-\bm{r}_3|) \nonumber\\
&&\times\,
D_{1}\left(t_1-t_2;\frac{\bm{r}_1+\bm{r}_4}{2},\frac{\bm{r}_2+\bm{r}_3}{2}\right)
\delta(t_1-t_2+t_3-t_4) \nonumber\\ && +\,2\pi N_1{\mathcal V}_1
w(|\bm{r}_1-\bm{r}_3|)w(|\bm{r}_2-\bm{r}_4|) \nonumber\\
&&\times\,
C_{1}\left(t_1-t_2;\frac{\bm{r}_1+\bm{r}_3}{2},\frac{\bm{r}_2+\bm{r}_4}{2}\right)
\delta(t_1-t_2+t_3-t_4)
 \nonumber\\ &&+ \langle
G^R_1(X_1,X_2)\rangle \langle G^A_1(X_3,X_4)\rangle   ,
\end{eqnarray}
where ${\mathcal V}_1$ is the volume of the first (left)
quantum dot,
$N_1$, $D_1(t,\bm{r},\bm{r}')$ and $C_1(t,\bm{r},\bm{r}')$
are respectively the density of states, the diffuson and the
Cooperon in the first (left) dot, $w(r)=e^{-r/2l}\sin k_Fr/ k_Fr$,
$k_F$ and $l$ are respectively the Fermi wave vector and elastic
mean free path. The same averaging procedure applies to the second
(right) dot. In the absence of magnetic field both the diffuson and the Cooperon
satisfy the same diffusion equation
\begin{eqnarray}
\left(\frac{\partial}{\partial t}-D\nabla^2\right)D_1(t,\bm{r},\bm{r}')=
{\mathcal V}_1\delta(t)\delta(\bm{r}-\bm{r}')
\end{eqnarray}
with the appropriate boundary conditions at the contacts.

Averaging the action (\ref{S2}) over disorder we will collect only
the terms proportional to the product of the Cooperons $C_1$ and
$C_2$ and ignore all other contributions which are unimportant for
the problem in question. In this way we arrive at the action
$S_{WL}$ which describes weak localization effects in our system.
Ignoring for simplicity the coordinate dependence of the Cooperons
$C_1,C_2$ inside the dots, we obtain
\begin{eqnarray}
&& iS_{WL}=-i\frac{g_2^2\delta_1\delta_2}{4\pi^2}\int dt_1\dots
dt_4 \int d\tau_1 d\tau_2\, {C}_{1}(t_1-\tau_1) \nonumber\\
&&\times\, {C}_{2}(t_2-\tau_2)
e^{i\big[\varphi^+(t_1)-\varphi^+(t_2)+\varphi^+(t_3)-\varphi^+(t_4)\big]}
\sin\frac{\varphi^-(t_1)}{2} \nonumber\\ &&\times\, \left[
h(\tau_1-t_2) e^{-i\frac{\varphi^-(t_2)}{2}}
 +  f(\tau_1-t_2) e^{i\frac{\varphi^-(t_2)}{2}} \right]
\nonumber\\ && \times\,
\left[
h(\tau_2-t_3)e^{i\frac{\varphi^-(t_3)}{2}}f(t_1+t_3-t_4-\tau_1)
\right.
\nonumber\\ &&
-\,\left. f(\tau_2-t_3)
e^{-i\frac{\varphi^-(t_3)}{2}}h(t_1+t_3-t_4-\tau_1)
\right]
\nonumber\\ &&\times\,
\left[ e^{-i\frac{\varphi^-(t_4)}{2}}f(-t_1+t_2+t_4-\tau_2)
+ e^{i\frac{\varphi^-(t_4)}{2}}\right.
\nonumber\\ &&
\times\, h(-t_1+t_2+t_4-\tau_2)\bigg]
+\,\big\{ 1\leftrightarrow 2, \varphi^\pm\to -\varphi^\pm\big\}.
\label{SWL}
\end{eqnarray}
Here we defined mean level spacing $\delta_{1,2}$ for both dots,
introduced ``classical'' $\varphi^+=(\varphi^F+\varphi^B)/2$ and
``quantum'' $\varphi^-=\varphi^F-\varphi^B$ phases and made use of
the Fourier transforms of the Fermi function $f(t)=\int
(dE/2\pi)\,f(E)e^{-iEt} \equiv\delta (t)-h(t)$. The resulting
action (\ref{SWL}) fully accounts for the effects of
electron-electron interactions on WL via the fluctuating phases
$\varphi^{\pm}$.

In order to find the WL correction to the current across the
central barrier $I_{WL}$ we employ the following general formula
\begin{equation}
I_{WL}=ie\int D^2\varphi^{\pm}\, \frac{\delta iS_{WL}[\varphi^{\pm}]}{\delta \varphi^-}\,
 e^{iS_{S}+iS^{(2)}_{2}}.
\label{I}
\end{equation}
The task at hand is to combine Eqs. (\ref{SWL}) and (\ref{I}) and
to average over the phases $\varphi^{\pm}$ by evaluating the path
integral in (\ref{I}). The contributions $S_C$ and $S_{\rm ext}$
in (\ref{Ss}) are quadratic in the fluctuating phases provided an
external circuit consists of linear elements. The remaining
contribution to $S_S$ (\ref{Ss}) which describes transfer of
electrons across the first and third barriers as well as AES
action of the second barrier $S^{(2)}_{2}$ are in general
non-Gaussian. However, in the interesting limit $g_{1,3} \gg 1$
phase fluctuations can be considered small down to exponentially
low energies \cite{PZ91,Naz99} in which case it suffices to expand
both contributions ${\rm tr}\,\ln[\check
G^{-1}]\big|_{t_2(\bm{r})=0}$ and $S^{(2)}_{2}$ up to the second
order $\varphi^{\pm}$. Furthermore, Gaussian approximation for the
first of these terms becomes essentially exact in the limit of
fully open outer barriers with $g_{1,3} \gg 1$ \cite{GGZ05}.

We conclude that the integral (\ref{I}) remains Gaussian at all
relevant energies and can easily be performed. After
straightforward algebra we arrive at the final result
\begin{eqnarray}
 && I_{WL} =\frac{e g_2^2\delta_1\delta_2}{8\pi^3}\,{\rm Re}\,
\int dEd\omega_1 d\omega_2 d\omega_3\, \tilde C_{2}(-\omega_2)
\nonumber\\ &&\times\, \tilde C_{1}(-\omega_3)
h(E-\omega_2)f(E+eV+\omega_3-\omega_1)
 \nonumber\\ &&\times\,
\big[ f(E+eV-\omega_1)h(E)\tilde P_1(\omega_1,\omega_2,\omega_3)
\nonumber\\ && +\,f(E+eV-\omega_1)f(E) \tilde
P_2(\omega_1,\omega_2,\omega_3) \nonumber \\ &&
+\,h(E+eV-\omega_1)h(E)\tilde P_2(\omega_1,\omega_3,\omega_2) \nonumber\\
&& +\,h(E+eV-\omega_1)f(E)\tilde P_3(\omega_1,\omega_2,\omega_3)
\big] \nonumber\\ && -\,\big\{ V\to -V \big\}. \label{IWL1}
\end{eqnarray}
Here $V$ is the average voltage across the central barrier,
$\tilde C_{1,2}$ are the Fourier transforms of the Cooperons
${C}_{1,2}(t)$ and the functions $\tilde P_{j}$ ($j=1,2,3$) are
defined as
\begin{eqnarray}
\tilde P_j(\omega_i)=\int\frac{dt_1dt_2dt_3}{(2\pi)^3}\,
e^{i[\omega_1t_1+\omega_2t_2+\omega_3t_3]} P_{j}(t_i),
\end{eqnarray}
where
\begin{equation}
P_{j}(t_1,t_2,t_3)=\exp [-\mathcal{F}(t_1,t_2,t_3)]{Q}_{j}(t_1,t_2,t_3), \label{calP}
\end{equation}
\begin{eqnarray}
&&\mathcal{F}(t_1,t_2,t_3) =F(t_1+t_3)+F(t_3)+F(t_1+t_2)\nonumber\\
&& +F(t_2)-F(t_1+t_2+t_3)-F(t_2-t_3),
\label{calF}
\end{eqnarray}
and $F(t)=\langle (\hat\varphi(t)-\hat\varphi (0))^2\rangle /2$
coincides with the phase correlation function (\ref{F}). The terms
${ Q}_j$ read
\begin{eqnarray}
&& { Q}_1=
e^{-i\left[K(t_2)+K(t_3)+K(|t_2-t_3|)\right]} \nonumber\\
&&\times\,
\big\{2e^{i\left[K(|t_1+t_2+t_3|)+K(t_1+t_3)+K(t_1+t_2)\right]}
\nonumber\\
&&-\,e^{i\left[K(t_1+t_2+t_3)+K(|t_1+t_3|)+K(|t_1+t_2|)\right]}
\big\},
\nonumber\\
&& {Q}_2=e^{i\left[K(|t_1+t_2+t_3|)-K(t_2)-K(|t_3|)\right]}
\nonumber\\ &&\times\,
e^{i\left[K(t_1+t_3)-K(|t_1+t_2|)-K(t_3-t_2)\right]},
\nonumber\\
&& {Q}_3= e^{i\left[K(t_1+t_2+t_3)-K(|t_2|)-K(|t_3|)\right]}
\nonumber\\ &&\times\,
e^{-i\left[K(t_1+t_3)+K(t_1+t_2)-K(|t_3-t_2|)\right]},
 \label{Pj}
\end{eqnarray}
where $K(t)=i\left\langle \big[\hat\varphi(0),\hat\varphi(t)\big]
\right\rangle$ is the response function. Eqs.
(\ref{IWL1})-(\ref{Pj}) fully determine WL correction to the
current in our system. The non-interacting result \cite{GZ06} is
reproduced by the first line of Eq. (\ref{IWL1}), while its second
and third lines {\it exactly} account for the effect of
interactions. We observe that the whole effect of
electron-electron interactions is encoded in two different
correlators of fluctuating phases $F(t)$ and $K(t)$ defined as
\begin{eqnarray}
F(t)&=&e^2\int\frac{d\omega}{2\pi}\,\omega\coth\frac{\omega}{2T}\,
{\rm Re}\big[Z(\omega)\big]\frac{1-\cos\omega t}{\omega^2},
\label{FFF}\\
K(t)&=&e^2\int\frac{d\omega}{2\pi}\,{\rm Re}\big[Z(\omega)\big]
\frac{\sin\omega t}{\omega}, \label{KKK}
\end{eqnarray}
where $Z(\omega)$ is an effective impedance ``seen'' by the
central barrier.

Let us recall that these correlation functions play a central in
the so-called $P(E)$-theory of Coulomb blockade in tunnel barriers
\cite{SZ,IN}. It turns out that exactly the same correlators
also describe the effect of electron-electron interactions on weak
localization. Below we will demonstrate that the phase correlation
function (\ref{FFF}) is responsible for electron dephasing
while the response function (\ref{KKK}) describes the
Coulomb blockade correction to WL.

\subsection{Dephasing}

Let us now turn to the analysis of the above general results. To
begin with, we notice that both functions (\ref{FFF}) and
(\ref{KKK}) are purely real and, hence, $|Q_j|\leq 1$.
Furthermore, at sufficiently long times $ |t| > \tau_{RC}$ (where
$\tau_{RC}$ is an effective $RC$-time of our system to be defined
later) we obtain
\begin{eqnarray}
F(t)& \simeq &\frac{2}{g_Z} \left(\ln\left|\frac{\sinh\pi Tt}{\pi
T\tau_{RC}}\right|+\gamma\right),
\label{Fas}\\
K(t)& \simeq & \frac{\pi}{g_Z}\,{\rm sign}\, t, \label{Kas}
\end{eqnarray}
where $g_Z = 2\pi /e^2 Z(0)=g_0+g_2$,
$g_0^{-1}=g_1^{-1}+g_3^{-1}+e^2R_S/2\pi$ and $\gamma \simeq 0.577$
is the Euler constant. The correlation function $F(t)$ grows with
time \cite{FN1} at any temperature including $T=0$. In contrast,
the response function $K(t)$ always remains small in the limit
$g_Z\gg 1$ considered here. Hence, the combination (\ref{calF})
should be fully kept in the exponent of (\ref{calP}) while the
correlator $K(t)$ can be safely ignored in the leading order in
$1/g_Z$. Then all ${ Q}_j \equiv 1$, the Fermi function $f(E)$
drops out from the result and we get $I_{WL}= G_{WL}V$, where
\begin{eqnarray}
 G_{WL}&=& -\frac{e^2 g_2^2\delta_1\delta_2}{8\pi^3}\int_0^\infty dt_2dt_3 \,
 e^{-\mathcal{F} (t_2,t_3)}
\nonumber\\ &&\times\,
{ C}_1(t_2){ C}_2(t_3),
\label{GWLF1}
\end{eqnarray}
where
\begin{equation}
\mathcal{F} (t_2,t_3)=2F(t_2)+2F(t_3)-F(t_2-t_3)-F(t_2+t_3).
\label{t2t3}
\end{equation}
Identifying $t_2=t_f-t$ and $t_3=t-t_i$ in Eq. (\ref{t2t3}) we
observe that the exponent $\exp (-\mathcal{F} (t_2,t_3))$ exactly
coincides with the expression (\ref{phase}) derived from simple
considerations involving electrons propagating along time-reversed
paths in an external fluctuating field. Thus, we arrive at an
important conclusion: In the leading order in $1/g_Z$ the WL
correction $G_{WL}$ is affected by electron-electron interactions
via dephasing produced {\it only} by the ``classical'' component
$\varphi^+$ of the fluctuating field which mediates such
interactions. This effect is described only by the phase
correlation function $F(t)$ (\ref{FFF}). At the same time,
fluctuations of the ``quantum'' field $\varphi^-$ turn out to be
irrelevant for dephasing and may only cause a (weak) Coulomb
blockade correction described by the response function $K(t)$
(\ref{KKK}). This latter effect will be analyzed in the next
subsection.

In order to simplify our consideration let us assume that two
quantum dots are identical, i.e. $g_{1}=g_3=g \gg 1, g_2$ and
$\delta_1=\delta_2=\delta$, and set $R_S \to 0$. In this case the
Cooperons are $C_1(t)=C_2(t)={e^{-t/\tau_D-t/\tau_H}}$, where
$\tau_D=4\pi /g\delta$ is the dwell time for each of the two dots
and $\tau_H \propto 1/H^2$ is the dephasing time due to the
external magnetic field $H$ which can be applied to the system. An
effective impedance $Z(\omega )$ takes the form
\begin{equation}
{\rm Re}\, Z(\omega)=\frac{e^2}{\pi g}\bigg[\frac{\tau^2}{\tau_{RC}^2}\frac{1}{1+\omega^2\tau^2}
+\frac{\pi\delta(\omega)}{\tau_D+\tau_{RC}}\bigg],
\label{imped}
\end{equation}
where $1/\tau =1/\tau_D+1/\tau_{RC}$, $\tau_{RC}=\pi /gE_C$,
$E_C=e^2/2(C+C_g+2C_J)$ and $C$, $C_J$ and $C_g$ are the
capacitances of respectively left(right) barriers, the central
junction and the gate electrode. Substituting the Cooperons
$C_{1,2}(t)$ into Eq. (\ref{GWLF}) we arrive at the final
expression for the WL correction $G_{WL} (T)$ in the presence of
electron-electron interactions
\begin{eqnarray}
G_{WL}= -\frac{e^2g^2\delta^2}{8\pi^3}\int_0^\infty dt_2 dt_3\,
e^{-\frac{(t_2+t_3)(\tau_D+\tau_H)}{\tau_D\tau_H}} \nonumber\\
\times\, e^{-\mathcal{F}(t_2,t_3)}. \label{GWLF}
\end{eqnarray}
With a good accuracy the double time integral in Eq. (\ref{GWLF})
can be replaced by a single one, in which case the
magnetoconductance $G_{WL}$ can be expressed in a much simpler
form
\begin{eqnarray}
\frac{G_{WL}}{G_{WL}^{(0)}}\simeq \frac{\tau_H}{\tau_D(\tau_H+\tau_D)}
\int_0^\infty dt\, e^{-2F(t/2)}e^{-t/\tau_D-t/\tau_H}.
\label{GWLsimple}
\end{eqnarray}

The result (\ref{GWLF}) for $H=0$ is plotted in Fig. 4. We
observe that electron-electron interactions always suppress
$G_{WL} (T)$ below its non-interacting value (\ref{G0}). This is a
direct consequence of interaction-induced electron dephasing
described by the correlation function $F(t)$.

\begin{figure}
\includegraphics[width=7cm]{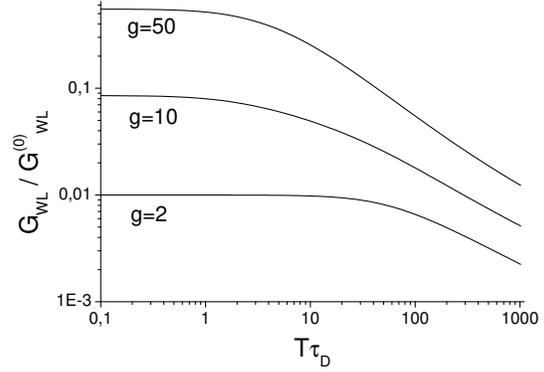}
\caption{Temperature dependence of WL correction $G_{WL}$
in the presence of electron-electron interactions for $\tau_D/\tau_{RC}=100$. }
\end{figure}

Let us define $u=\tau_D/\tau_{RC}=4E_C/\delta$
and consider the limit of metallic
dots $u\gg 1$. At $g/\tau_D\lesssim T\lesssim
1/\tau_{RC}$ and for $g\gtrsim 8$ we find
\begin{equation}
\frac{G_{WL}}{|G_{WL}^{(0)}|}\simeq 
-\left(\frac{g}{4}-2\gamma\right)
\frac{\left({2\pi}/{u}\right)^{{8}/{g}}}{(\pi T\tau_D)^{1-8/g}},
\label{tT}
\end{equation}
whereas at $T\tau_D \lesssim 1$ the WL correction saturates to
\begin{eqnarray}
{G_{WL}}/{|G_{WL}^{(0)}|}&\simeq & -\left(2/u\right)^{{8}/{g}},\;\; g\gtrsim 8,
\nonumber\\
{G_{WL}}/{|G_{WL}^{(0)}|}&\simeq & -g/2u,\;\; 1\lesssim g\lesssim 8,
\label{t0}
\end{eqnarray}

Let us compare the magnitude of the WL correction (\ref{t0}) to that of the
leading contribution to the system conductance evaluated in the presence of
electron-electron interactions. This contribution reads \cite{PZ88,SZ,IN}
\begin{equation}
\frac{dI(T=0)}{dV}=\frac{1}{R_2}
\frac{\left(e^{-\gamma}eV\tau_{RC}\right)^{4/g}}{\Gamma\left(1+{4}/{g}\right)},
\label{main}
\end{equation}
where $\Gamma (x)$ is the gamma-function.
Setting $eV\sim 1/\tau_D$ in the above expression,
we observe that the WL correction (\ref{t0}) remains much smaller
than Eq. (\ref{main}) by the parameter $\sim g_2/g_1g_3u^{4/g}$.

\begin{figure}
\includegraphics[width=6.5cm]{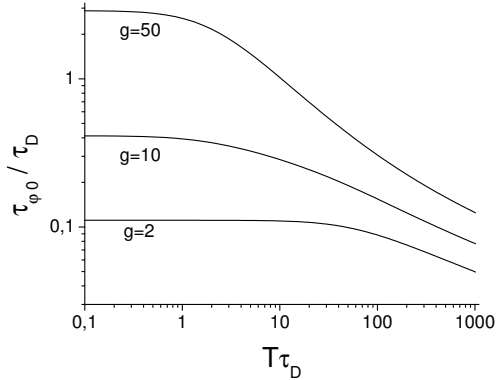}
\caption{Temperature dependence of the electron
dephasing time $\tau_{\varphi}$  in the presence of interactions
for $\tau_D/\tau_{RC}=100$. }
\end{figure}

Let us phenomenologically define the electron decoherence time
$\tau_\varphi$ by taking the Cooperons (for $H \to 0$) in the form
$C_{1,2}(t)={e^{-t/\tau_D-t/\tau_\varphi}}$. This definition obviously
yields \cite{GZ06} ${G_{WL}}/{G_{WL}^{(0)}}=(1+\tau_D/\tau_\varphi)^{-2}$.
Resolving this equation with respect to $\tau_\varphi$, we obtain
\begin{equation}
\tau_\varphi /\tau_D=\left(\sqrt{G_{WL}^{(0)}/G_{WL}}-1\right)^{-1}.
\label{tphi}
\end{equation}
Substituting the result (\ref{tT}) into Eq. (\ref{tphi}) at
sufficiently high temperatures $g/\tau_D\lesssim T\lesssim
1/\tau_{RC}$ and in the limit $g\gg 1$ we obtain
\begin{equation}
\tau_\varphi (T)\sim
\sqrt{g}\tau_D\left(\frac{\delta}{E_C}\right)^{4/g}(T\tau_D)^{4/g-1/2}.
\end{equation}
Combining Eqs. (\ref{t0}) and (\ref{tphi}) at lower temperatures
$T\tau_D \lesssim 1$ and for $g \gtrsim 8$ we find
\begin{equation}
\tau_\varphi (T)=\tau_{\varphi 0}=\frac{\tau_D}{(2E_C/\delta
)^{4/g}-1}.
\label{tauphi0}
\end{equation}
In the limit of large $g \gg 1$ this expression yields
\begin{equation}
\tau_{\varphi 0}\simeq \frac{g\tau_D}{4\ln(2E_C/\delta)}
=\frac{\pi}{\delta \ln(2E_C/\delta)}. \label{tauphi}
\end{equation}

The results for the electron decoherence time $\tau_\varphi (T)$
(\ref{tphi}) are also plotted in Fig. 5 for different values of
$g$. We observe that at higher temperatures $\tau_\varphi (T)$
shows a $g$-dependent power law dependence on $T$ and eventually
saturates to the value (\ref{tauphi}) at lower $T$. This is
exactly the behavior observed in a number of experiments with
quantum dots \cite{Bird1,Clarke,Pivin,Huibers,Hackens}. We will
postpone a detailed comparison between our theory and experiments
to Sec. 6.

\subsection{Perturbation theory and Coulomb correction to weak localization}

Let us now analyze the role of the response function $K(t)$ which
was disregarded in the previous subsection. For this purpose let
us expand the general expression for the current (\ref{IWL1}) to
the first order in the interaction, i.e. to the first order in
both $F(t)$ and $K(t)$. We get
\begin{equation}
I_{WL}=G_{WL}^{(0)}V+\delta I_{WL}^{F}(V)+\delta I_{WL}^K(V),
\label{pert}
\end{equation}
where
\begin{eqnarray}
&& \delta I_{WL}^F=-\frac{e^4g_2^2\delta_1\delta_2 V}{8\pi^3}
\int\frac{d\omega}{2\pi}\,\frac{{\rm Re}\,Z(\omega)}{\omega}
\coth\frac{\omega}{2T}
\nonumber\\ &&\times\,
\big[2\tilde C_1(0)\tilde C_2(\omega)+2\tilde C_1(\omega)\tilde C_2(0)
-2\tilde C_1(0)\tilde C_2(0)
\nonumber\\ &&
-\,\tilde C_1(\omega)\tilde C_2(\omega)-\tilde C_1(-\omega)\tilde C_2(\omega)
\big],
\label{WLF}
\end{eqnarray}
\begin{eqnarray}
&& \delta I_{WL}^K=\frac{e^3 g_2^2\delta_1\delta_2}{16\pi^3}
\int\frac{d\omega}{2\pi}\frac{W(\omega,V)}{\omega}\big\{
\,{\rm Re}\,Z(\omega)
\nonumber\\ &&\times\,
\big[ 2\tilde C_1(0)\tilde C_2(\omega)
+2\tilde C_1(\omega)\tilde C_2(0) - \tilde C_1(-\omega)\tilde C_2(\omega)\big]
\nonumber\\ &&
+\, i\,{\rm Im}\, Z(\omega)\; \tilde C_1(\omega)\tilde C_2(\omega)
\big\}.
\label{WLK}
\end{eqnarray}
Here we defined the function
$$
W=(\omega+eV)\coth\frac{\omega+eV}{2T}
-(\omega-eV)\coth\frac{\omega-eV}{2T}.
$$

The above expressions allow to make several important observations.
To begin with, we note that the term $\delta I_{WL}^F$
is linear in the bias voltage while $\delta I_{WL}^K$ is non-linear in $V$.
The physical meaning of these two terms is entirely different. While,
as we already know, the correction  $\delta I_{WL}^F$ describes electron
dephasing, the term $\delta I_{WL}^K$ is nothing but the {\it Coulomb
blockade correction} to $G_{WL}$ (\ref{GWLF}). The latter conclusion
is supported, for instance, by the observation that at
large voltages $\delta I_{WL}^K$ tends to a
constant offset value, which is a typical sign of Coulomb blockade.

Just for the sake of illustration, let us for a moment assume that
the environment remains Gaussian at all energies. In this case
$dI_{WL}^K/dV$ diverges at $V,T \to 0$ indicating the importance
of higher order perturbative terms in the low energy limit. Making
use of analytical properties of the functions $\tilde
P_j(\omega_1,\omega_2,\omega_3)$ one can exactly sum up diverging
perturbative series to all orders and quite generally demonstrate
that in the limit of zero voltage and temperature the WL
correction tends to zero, $dI_{WL}/dV\to 0$, implying complete
Coulomb suppression of weak localization in this limit. This
strong Coulomb blockade of WL can be recovered only if one fully
accounts for the response function $K(t)$. We also note that at
large conductances weak Coulomb blockade may turn into strong one
only at exponentially small energies \cite{PZ91,Naz99} typically
well below the inverse electron dwell time $1/\tau_D$, otherwise
Coulomb suppression of WL remains weak.

Turning back to the first order terms (\ref{WLF}) and (\ref{WLK}),
we observe that in the linear response regime $V\to 0$ almost all
contributions from $\delta I_{WL}^F$ and $\delta I_{WL}^K$ cancel
each other exactly in the limit $T \to 0$. This (partial)
cancellation of the so-called ``coth''(\ref{WLF}) and ``tanh''
(\ref{WLK}) terms is a general feature of the first order
perturbation theory. For instance, in diffusive conductors it has
been observed by various authors \cite{FA,AAG}. This cancellation
is sometimes interpreted as a ``proof'' of zero dephasing of
interacting electrons at $T=0$ {\it assuming} that the same
``coth-tanh'' combination should occur in every order of the
perturbation theory \cite{FN2}. Our exact result (\ref{IWL1})
demonstrates, that this assumption is incorrect. Partial
cancellation between dephasing and Coulomb blockade terms occurs
only in the first order and is of little importance for the issue
of electron dephasing. Actually, the combination $1-2f(E)\equiv
\tanh (E/2T)$ enters only in the first order, while the exact
expression (\ref{IWL1}) depends on $f(E)$ and $1-f(E)$ and does
not contain the combination $1-2f(E)$ at all. This observation is
fully consistent with our general result for the WL correction
\cite{GZ03} expressed in terms of the matrix elements of the
operators $\hat \rho$ and $1-\hat \rho$, where $\rho$ is the
electron density matrix. In fact, the analysis presented above is
a way to explicitly evaluate these matrix elements for a
particular case of two quantum dots.

Finally, let us estimate the corrections $\delta I_{WL}^F$ and
$\delta I_{WL}^K$. For max$(eV,T) \gtrsim 1/\tau_D$ we obtain
\begin{eqnarray}
&&\frac{dI^F_{WL}}{dV}=\frac{8e^2 g_2^2}{\pi g^3}
\left[\pi T\tau_D +2\ln\frac{4E_C}{\delta}-1 \right],
\label{dIF}\\
&& \frac{dI^K_{WL}}{dV}=\left\{
\begin{array}{cc}
\frac{20}{3}\frac{e^2g_2^2}{g^3 T\tau_D}, & eV\ll T, \\
\frac{6}{\pi}\frac{e^2g_2^2}{g^3(eV\tau_D)^2}, & eV \gg T,
\end{array}
\right.
\label{dIK}
\end{eqnarray}
i.e. at $T\tau_D\sim 1$ the ratio of the terms (\ref{dIK}) and
(\ref{dIF}) is  of order $\sim 1/\ln (4E_C/\delta )\ll 1$ and it
decreases further with increasing temperature as $\propto
1/T^2\tau_D^2$. This estimate demonstrates that ${dI^K_{WL}}/{dV}$
can be ignored as compared to the main contribution
${dI^F_{WL}}/{dV}$. This conclusion remains valid also beyond the
first order perturbation theory. Indeed, keeping the function
$F(t)$ in the exponent of Eq. (\ref{IWL1}) and expanding this
result only in powers of $K(t)$, at $T\tau_D\gtrsim 1$ one finds
\begin{equation}
\frac{dI^K_{WL}}{dV}\sim \frac{G_{WL}}{gT\tau_D}\ll G_{WL},
\end{equation}
thus supporting our conclusion about relative unimportance of the
Coulomb blockade correction to $G_{WL}$ for the problem in
question.

\subsection{Summary of approximations}

For clarity, let us again summarize all the approximations used in
the above analysis:

$(i)$ Throughout our calculation we have used the equilibrium form
of the distribution function matrices $\check F_{1,2}$
(\ref{distr}) which effectively implies neglecting the dependence
of the Green functions $\check G_{1,2}$ on the phases $\varphi_1$
and $\varphi_3$. This is accurate except at energies well below
the inverse dwell time $1/\tau_D$. The WL correction
(\ref{GWLF1})-(\ref{t2t3}) saturates at energies $\sim 1/\tau_D$
and, hence, is totally insensitive to this approximation. The
Coulomb term $dI^K_{WL}/dV$ should be treated somewhat more
carefully in the limit $T,eV\ll 1/\tau_D$. For $R_S \to 0$ this
treatment also yields {\it saturation} of the Coulomb correction
to WL at energies of order $1/\tau_D$, exactly as in the case of
the interaction correction to the Drude conductance
\cite{GZ041,GZ042}. In other words, at all $T,eV < 1/\tau_D$ the
term $dI^K_{WL}/dV$ remains constant of order $\sim G_{WL}/g \ll
G_{WL}$. Thus, the above approximation is completely unimportant
for any of our conclusions.

$(ii)$ We have assumed $g_{1,3} \gg 1$ and $g_{1,3} \gg g_2$. The
first inequality just implies that our structure is metallic,
while the second one is only a matter of technical convenience.
Obviously, our results for $\tau_\varphi$ remain qualitatively the
same also for $g_{1,3} \sim g_2$.

$(iii)$ For completeness, let us also mention that in the final
expressions for $G_{WL}$, e.g., in Eqs. (\ref{tT}) and (\ref{t0}),
we have disregarded small renormalization of the conductance $g$
due to Coulomb effects \cite{GZ01,BN}. This renormalization is
strictly zero for fully open outer barriers, otherwise it can be
trivially included into our final results.

Thus, our treatment presented in this section does not contain any
approximations which could influence our main results and
conclusions. In the limit $(ii)$ our general results for the WL
correction to the $I-V$ curve in the presence of electron-electron
interactions (\ref{IWL1})-(\ref{Pj}) are exact.

\section{Arrays of quantum dots and diffusive conductors}

One of the main conclusions reached in the previous section is
that the electron decoherence time is fully determined by
fluctuations of the phase fields $\varphi^+$ (and the correlation
function $F(t)$), whereas the phases $\varphi^-$ (and the response
function $K(t)$) are irrelevant for $\tau_\varphi$ causing only a
weak Coulomb correction to $G_{WL}$. This conclusion is general
being independent of a number of scatterers in our system. Note
that exactly the same conclusion was already reached in the case
of diffusive metals by means of a different approach
\cite{GZ99,GZ00,GZ03}. Thus, in order to evaluate the decoherence
time for interacting electrons in arrays of quantum dots it is
sufficient to account for the fluctuating fields $V^+$ totally
ignoring the fields $V^-$. The corresponding calculation is
presented below. We will specifically address the example of 1d
arrays and argue that the final result for the zero temperature
decoherence time $\tau_{\varphi 0}$ actually does not depend on
the dimensionality of the array. We will also demonstrate how our
present results for $\tau_{\varphi}$ match with those obtained
previously for weakly disordered metals \cite{GZ98,GZ99,GZ00}.

\subsection{1d structures}

Let us consider a 1D array of $N-1$ quantum dots by $N$ identical
barriers as shown in Fig. 3. For simplicity, we will stick to the
case of identical barriers (with dimensionless conductance $g \gg
1$ and Fano factor $\beta$) and identical quantum dots (with mean
level spacing $\delta$ and dwell time $\tau_D=2\pi/\delta g$). The
WL correction to the system conductance has the form \cite{GZ06}
\begin{eqnarray}
 G_{WL}&=&-\frac{e^2g\delta}{4\pi^2N^2}\sum_{n=1}^N
\int_0^\infty dt\, \nonumber\\ &&\times\, \big\{
\beta\big[C_{n-1,n}(t)+C_{n,n-1}(t)\big] \nonumber\\ &&
+\,(1-\beta)\big[C_{nn}(t)+C_{n-1,n-1}(t)\big] \big\}. \label{GWL}
\end{eqnarray}
The Cooperon $C_{nm}(t)$ is determined from a discrete version of
the diffusion equation. For non-interacting electrons and in the
absence of the magnetic field this equation reads
\begin{eqnarray}
\frac{\partial C_{nm}}{\partial
t}+\frac{2C_{nm}-C_{n-1,m}-C_{n+1,m}}{2\tau_D}=\delta_{nm}\delta(t).
\label{EqC}
\end{eqnarray}
The boundary conditions for this equation are $C_{nm}=0$ as long
as the index $n$ or $m$ belongs to one of the bulk electrode. The
solution of Eq. (\ref{EqC}) with these boundary conditions can
easily be obtained. We have
\begin{eqnarray}
C_{n,m}^{(0)}(t)=\frac{2}{N}\sum_{q=1}^{N-1}\int\frac{d\omega}{2\pi}\,e^{-i\omega
t}\,\frac{\sin\frac{\pi qn}{N}\sin\frac{\pi qm}{N}}
{-i\omega+\frac{1-\cos\frac{\pi q}{N}}{\tau_D}}. \label{C0}
\end{eqnarray}
This solution can be represented in the form
$C^{(0)}_{nm}(t)=C^{\rm bulk}_{n-m}(t)-C^{\rm bulk}_{n+m}(t)$,
where
\begin{eqnarray}
C_{n-m}^{\rm
bulk}(t)=\frac{1}{N}\sum_{q=1}^{N-1}\int\frac{d\omega}{2\pi}\,e^{-i\omega
t}\,\frac{\cos\frac{\pi q(n-m)}{N}}
{-i\omega+\frac{1-\cos\frac{\pi q}{N}}{\tau_D}}. \label{Cbulk}
\end{eqnarray}

In the limit of large $N$ the term $C^{\rm bulk}_{n+m}(t)$ can be
safely ignored and we obtain $C_{nm}(t)\approx C^{\rm
bulk}_{n-m}(t)$. Let us express the contribution $C^{\rm
bulk}_{n-m}(t)$ as a sum over the integer valued paths
$\nu(\tau)$, which start in the $m-$th dot and end in the $n-$th
one (i.e. $\nu(0)=m$, $\nu(t)=n$) jumping from one dot to another
at times $t_j$. This expression  can be recovered if one expands
Eq. (\ref{Cbulk}) in powers of $\tau_D^{-1}\cos[{\pi q}/{N}]$ with
subsequent summation over $q$ in every order of this expansion.
Including additional phase factors acquired by electrons in the
presence of the fluctuating fields $V_\nu^+$, we obtain
\begin{eqnarray}
&&
C_{nm}(t)=\sum_{k=|n-m|}^\infty\;{\sum_{\nu(\tau)}}\bigg|_{\nu(0)=m}^{\nu(t)=n}
\nonumber\\ &&\times\, \frac{1}{(2\tau_D)^k}\int_0^t
dt_k\int_0^{t_k}dt_{k-1}\dots \int_0^{t_3}dt_2\int_0^{t_2}dt_1
\nonumber\\ &&\times\,
e^{-\frac{t-t_k}{\tau_D}}e^{-\frac{t_k-t_{k-1}}{\tau_D}}\dots
e^{-\frac{t_2-t_2}{\tau_D}}e^{-\frac{t_2-t_1}{\tau_D}}e^{-\frac{t_1}{\tau_D}}
\nonumber\\ &&\times\, \exp\left\{i\int_0^t
d\tau\big[eV^+_{\nu(\tau)}(\tau)-eV^+_{\nu(t-\tau)}(\tau)\big]\right\}.
\end{eqnarray}
Averaging over Gaussian fluctuations of voltages $V^+$ and
utilizing the symmetry of the voltage correlator $ \langle
V_{\nu_1}^+(\tau_1) V_{\nu_2}^+(\tau_2)\rangle =\langle
V_{\nu_2}^+(\tau_1) V_{\nu_1}^+(\tau_2)\rangle$, we get
\begin{eqnarray}
&&
C_{nm}(t)=\sum_{k=|n-m|}^\infty\;{\sum_{\nu(\tau)}}\bigg|_{\nu(0)=m}^{\nu(t)=n}
\nonumber\\ &&\times\, \frac{e^{-t/\tau_D}}{(2\tau_D)^k}\int_0^t
dt_k\int_0^{t_k}dt_{k-1}\dots \int_0^{t_3}dt_2\int_0^{t_2}dt_1
\nonumber\\ &&\times\, \exp\bigg\{-e^2\int_0^t d\tau_1\int_0^t
d\tau_2 \big[ \langle V_{\nu(\tau_1)}^+(\tau_1)
V_{\nu(\tau_2)}^+(\tau_2)\rangle \nonumber\\ && -\,\langle
V_{\nu(\tau_1)}^+(\tau_1) V_{\nu(t-\tau_2)}^+(\tau_2)\rangle \big]
\bigg\}. \label{C2}
\end{eqnarray}

The correlator of voltages can be obtained, e.g., with the aid of
the $\sigma$-model approach employed in Ref. \cite{GZ06}.
Integrating over Gaussian fluctuations of the $Q$-fields one
arrives at the quadratic action for the fluctuating fields $V^+$
which has the form
\begin{eqnarray}
&& iS=\frac{2i}{N}\sum_{q=1}^N\int\frac{d\omega}{2\pi}
\bigg[4C\left(1-\cos\frac{\pi q}{N}\right)+C_g \nonumber\\ &&
+\,\frac{g\tau_De^2}{\pi}\frac{1-\cos\frac{\pi q}{N}}
{-i\omega\tau_D+1-\cos\frac{\pi q}{N}}
\bigg]V^+_q(\omega)V^-_q(-\omega) \nonumber\\ &&
-\,\frac{2}{N}\sum_{q=1}^N\int\frac{d\omega}{2\pi}
\frac{g\tau_D^2e^2}{\pi} \frac{\left(1-\cos\frac{\pi
q}{N}\right)\omega\coth\frac{\omega}{2T}}
{\omega^2\tau_D^2+\left(1-\cos\frac{\pi q}{N}\right)^2}
\nonumber\\ &&\times\, V^-_q(\omega)V^-_q(-\omega). \label{S11}
\end{eqnarray}
Here we defined
\begin{equation}
V_q^\pm(\omega)=\sum_{n=1}^{N-1}\int dt\; \sin\frac{\pi q}{N}\,
e^{i\omega t}\, V_n^\pm(t).
\end{equation}
The action (\ref{S11}) determines the expressions for both
correlators $\langle V^+V^+\rangle$ ($F$-function) and $\langle
V^+V^-\rangle$ ($K$-function) responsible respectively for
decoherence and Coulomb blockade correction to WL. Since our aim
is to describe electron decoherence, only the first out of these
two correlation functions is of importance for us here. It reads
\begin{eqnarray}
&& \langle
V^+_n(t_1)V^+_m(t_2)\rangle=\frac{2}{N}\sum_{q=1}^{N-1}\int\frac{d\omega}{2\pi}\,
e^{-i\omega(t_1-t_2)} \nonumber\\ &&\times\, \frac{\frac{g
e^2}{\pi}\left(1-\cos\frac{\pi q}{N}\right)\, \sin\frac{\pi
qn}{N}\sin\frac{\pi qm}{N}} {\left| 4C\left(1-\cos\frac{\pi
q}{N}\right)+C_g +\frac{g\tau_De^2}{\pi}\frac{1-\cos\frac{\pi
q}{N}} {-i\omega\tau_D+1-\cos\frac{\pi q}{N}} \right|^2}
\nonumber\\ &&\times\, \frac{\tau_D^2\,\omega
\coth\frac{\omega}{2T}} {\omega^2\tau_D^2+\left( 1-\cos\frac{\pi
q}{N}\right)^2}. \label{VV}
\end{eqnarray}
In the continuous limit $N \gg 1$ and for sufficiently low
frequencies $\omega \ll 1/\tau_D$ both correlators $\langle
V^+V^+\rangle$ and $\langle V^+V^-\rangle$ defined by Eq.
(\ref{S11}) reduce to those of a diffusive metal \cite{GZ99}.

To proceed let us consider diffusive paths $\nu(\tau)$, in which
case one has
\begin{eqnarray}
\langle V_{\nu(\tau_1)}^+(\tau_1) V_{\nu(\tau_2)}^+(\tau_2)\rangle
&\approx & \frac{1}{N-1}\sum_{n,m=1}^{N-1}\langle V_{n}^+(\tau_1)
V_{m}^+(\tau_2)\rangle \nonumber\\ &&\times\,
D_{nm}(|\tau_1-\tau_2|),
\label{VVD}
\end{eqnarray}
where $D_{nm}(\tau)$ is the diffuson. For $H \to 0$ it exactly
coincides with the Cooperon for non-interacting electrons
(\ref{C0}), $D_{nm}(t)=C_{n,m}^{(0)}(t)$, i.e.
\begin{eqnarray}
D_{nm}(t)=\frac{2}{N}\sum_{q=1}^{N-1}\int\frac{d\omega}{2\pi}\,e^{-i\omega
t}\,\frac{\sin\frac{\pi qn}{N}\sin\frac{\pi qm}{N}}
{-i\omega+\frac{1-\cos\frac{\pi q}{N}}{\tau_D}}. \label{DC0}
\end{eqnarray}
Substituting Eq. (\ref{VVD}) into (\ref{C2}), we obtain
\begin{equation}
C_{nm}(t)\approx C^{(0)}_{nm}(t)\, e^{-\mathcal{F}(t)}, \label{C3}
\end{equation}
where
\begin{eqnarray}
\mathcal{F}(t)&=&\frac{e^2}{N-1}\sum_{n,m=1}^{N-1}\int_{0}^t
dt_1dt_2 \langle V_n^+(t_1)V_m^+(t_2) \rangle \nonumber\\
&&\times\, \big[ D_{nm}(|t_1-t_2|)-D_{nm}(|t-t_1-t_2|)\big]
\label{F3}
\end{eqnarray}
is the function which controls the Cooperon decay  in time, i.e.
describes electron decoherence for our 1d array of quantum dots.
The WL correction $G_{WL}$ in the presence of electron-electron
interactions is recovered by substituting the result (\ref{C3})
into Eq. (\ref{GWL}).

Note that in the limit of very large $N$ and for $t\gg \tau_D$ Eq.
(\ref{F3}) (combined with (\ref{VV}) and (\ref{DC0})) reduces to
Eq. (22) of Ref. \cite{GZ00} derived for metallic conductors by
means of a different technique. Replacing the sum in (\ref{F3}) by
momentum integration, in the same limit one arrives at the result
which matches with Eq. (23) of \cite{GZ00} provided one sets the
charging energy terms equal to zero.

Since the behavior of the latter formula was already analyzed in
details earlier \cite{GZ00}, there is no need to repeat this
analysis here. The dephasing time $\tau_\varphi$ can be extracted
from the equation $\mathcal{F}(\tau_\varphi)=1$. From Eq.
(\ref{F3}) with a good accuracy we obtain
\begin{eqnarray}
\frac{1}{\tau_\varphi}=\frac{e^2}{N-1}\sum_{n,m=1}^{N-1}\int d\tau
\langle V_n^+(\tau)V_m^+(0) \rangle D_{nm}(\tau). \label{tau11}
\end{eqnarray}
Combining this formula with Eqs. (\ref{VV}) and (\ref{DC0}), in
the most interesting limit $T\to 0$ and for $\tau_D\gg R(4C+C_g)$
we find
\begin{eqnarray}
\frac{1}{\tau_{\varphi 0}}&=&\frac{1}{2g\tau_D(N-1)}
\sum_{q=1}^{N-1}
\ln\frac{2e^2}{\delta\left(4C\left(1-\cos\frac{\pi
q}{N}\right)+C_g\right)}, \nonumber
\end{eqnarray}
which yields
\begin{equation}
\tau_{\varphi 0}=\frac{2g\tau_D}{\ln(4\tilde E_C/\delta)}
=\frac{4\pi}{\delta\ln(4\tilde E_C/\delta)}, \label{t1}
\end{equation}
where $\tilde E_C=e^2/2C_g$ for $C_g\gg C$ and  $\tilde
E_C=e^2/4C$ in the opposite case $C_g\ll C$.

We observe that apart from an unimportant numerical factor of
order one the result for $\tau_{\varphi 0}$ (\ref{t1}) derived for
1d array of quantum dots coincides with the exact result
(\ref{tauphi}) derived in the previous section for the case of two
quantum dots. Thus, we arrive at an important conclusion: In the
low temperature limit the electron decoherence time $\tau_{\varphi
0}$ is practically independent of the number of scatterers in the
conductor (provided the latter number exceeds two) and is
essentially determined by {\it local} properties of the system.

In order to determine the dephasing length $L_\varphi
=\sqrt{D\tau_\varphi}$ let us define the diffusion coefficient
\begin{equation}
D=\frac{d^2}{2\tau_D}=\frac{d^2 g\delta}{4\pi}, \label{DDD}
\end{equation}
where $d\equiv \mathcal{V}^{1/3}$ is the average dot size.
Combining Eqs. (\ref{t1}) and (\ref{DDD}), at $T=0$ we obtain
\begin{equation}
L_{\varphi 0}= \sqrt{D\tau_{\varphi 0}}= d\sqrt{g/\ln (4\tilde
E_C/\delta )}.
\end{equation}

At non-zero $T$ thermal fluctuations provide an additional
contribution to the dephasing rate $1/\tau_\varphi$. Again
substituting Eqs. (\ref{VV}) and (\ref{DC0}) into (\ref{tau11}),
we get
\begin{equation}
\frac{1}{\tau_\varphi(T)}\simeq \frac{1}{\tau_{\varphi
0}}+\frac{\pi T}{3 g}\min\{ N,N_\varphi \}, \label{tauT}
\end{equation}
where $N_\varphi =L_\varphi /d \sim \sqrt{\tau_\varphi/\tau_D}$ is
the number of quantum dots within the length $L_\varphi$. We
observe that for sufficiently small $N<N_\varphi$ (but still $N\gg
1$) the dephasing rate increases linearly both with temperature
and with the number $N$. At larger $N>\sqrt{g/\ln[{4\tilde
E_C}/{\delta}]}$ and/or at high enough temperatures $N_\varphi$
becomes smaller than $N$ and Eq. (\ref{tauT}) for $\tau_\varphi$
should be resolved self-consistently. In this case we obtain
\begin{equation}
\tau_\varphi\simeq (3g\sqrt{\tau_D}/\pi T)^{2/3}, \label{AAK}
\end{equation}
which matches with the AAK result \cite{AAK}. Eq. (\ref{tauT})
also allows to estimate the temperature $ T^*\simeq 3
g/[\pi\tau_{\varphi 0}\min\{N,N_\varphi\}]$ at which the crossover to
the temperature-independent regime (\ref{t1}) occurs. We find
\begin{eqnarray}
&& T^*\simeq \frac{3\ln[{4\tilde E_C}/{\delta}]}{2\pi N\tau_D},\;\;\;
 N\lesssim \sqrt{\frac{g}{\ln[{4\tilde E_C}/{\delta}]}},
\nonumber\\ && T^*\simeq \frac{3\ln^{3/2}[{4\tilde
E_C}/{\delta}]}{2\pi \tau_D\sqrt{g}},\;\;\;
 N\gtrsim \sqrt{\frac{g}{\ln[{4\tilde E_C}/{\delta}]}}.
\end{eqnarray}

\subsection{Good metals and granular conductors}

The above analysis and conclusions can be generalized further to
the case 2d and 3d structures. This generalization is absolutely
straightforward (see, e.g. \cite{GZ06}) and therefore is not
presented here. At $T \to 0$ one again arrives at the same result
for $\tau_{\varphi 0}$ (\ref{t1}).

Now we discuss the relation between our present results and those
derived earlier for weakly disordered metals by means of a
different approach \cite{GZ98,GZ99,GZ00}. Let us express the dot
mean level spacing via the average dot size $d$ as $\delta=
1/N_0d^3$ (where $N_0=mp_F/2\pi^2$ is the electron density of
states at the Fermi level). Then we obtain
\begin{equation}
D=\frac{g}{4\pi N_0 d}.
\label{DDDD}
\end{equation}
Below we consider two different physical limits of $(a)$ good
metals and $(b)$ strongly disordered (granular) conductors. For
the model $(a)$ we assume that quantum dots are in a good contact
with each other. In this case $g$ scales linearly with the contact
area $\mathcal{A}=\gamma d^2$, where $\gamma$ is a numerical
factor of order (typically smaller than) one which particular
value depends on geometry. For weakly disordered metals most
conducting channels in such contacts can be considered open.
Hence, $g=p_F^2\mathcal{A}/2\pi$ and
\begin{equation}
D=\gamma v_Fd/4, \label{vFd}
\end{equation}
i.e. $D \propto d$. Comparing this estimate with the standard
definition of $D$ for a bulk diffusive conductor, $D=v_Fl/3$, we
immediately observe that within our model the average dot size is
comparable to the elastic mean free path, $l \sim \gamma d$, as it
should be for weakly disordered metals.

Expressing $\tau_{\varphi 0}$ (\ref{t1}) via $D$, in this limit we
get
\begin{equation}
\tau_{\varphi 0}=\frac{64}{\pi
\gamma^3}\frac{m^2}{v_F^2}\frac{D^3}{\ln (D/D_{c1})}, \label{D3}
\end{equation}
where $m$ is the electron mass and $D_{c1}$ is constant which
depends on $\tilde E_C$. Estimating, e.g., $\tilde E_C \approx
e^2/2d$,  one obtains $D_{c1}^{-1}=4\pi\sqrt{2}eN_0^{3/2}$.

Note that apart from an unimportant numerical pre-factor and the
logarithm in the denominator of Eq. (\ref{D3}) the latter result
for $\tau_{\varphi 0}$ coincides with that derived in Refs.
\cite{GZ98,GZ99,GZ00} for a bulk diffusive metal within the
framework of a completely different approach, cf., e.g., Eq. (81) in
\cite{GZ99}. Within that approach local properties of the model
could not be fully defined. For this reason in the corresponding
integrals in \cite{GZ98,GZ99,GZ00} we could not avoid using an
effective high frequency cutoff procedure which yields the correct
leading dependence $\tau_{\varphi 0} \propto D^3$ and it only does
not allow to recover an additional logarithmic dependence on $D$
in (\ref{D3}). Our present approach is divergence-free and, hence,
it does not require any cutoffs.

We can also add that Eq. (\ref{t1}) also agrees with our earlier
results \cite{GZ98,GZ99,GZ00} derived for quasi-1d and quasi-2d
metallic conductors. Provided the 
transversal size $a$ of our array is
smaller than $d$ one should set $\mathcal{A} \sim da$ for 2d and
$\mathcal{A} \sim a^2$ for 1d conductors.
Then Eq. (\ref{t1}) yields  $\tau_{\varphi 0} \propto D^2/\ln D$ 
and  $\tau_{\varphi 0} \propto D/\ln D$  respectively in 2d and 1d cases.
Up to the factor $\ln D$ these dependencies coincide with ones derived
previously, cf., e.g., Eq. (32) in \cite{GZ00}.

Now let us turn to the model $(b)$ of strongly disordered or
granular conductors. In contrast to the situation $(a)$, we will
assume that the contact between dots (grains) is rather poor, and
inter-grain electron transport may occur only via a limited number
of conducting channels. In this case the average dimensionless
conductance $g$ can be approximated by some
$\mathcal{A}$-independent constant $g=g_c$. Substituting $g_c$
instead of $g$ into Eq. (\ref{DDDD}) we observe that in the case
of strongly disordered structures one can expect $D \propto 1/d$.
Accordingly, for $\tau_{\varphi 0}$ (\ref{t1}) one finds
\begin{equation}
\tau_{\varphi 0}=\frac{g_c^3}{32\pi^2 N_0^2D^3\ln (D_{c2}/D)},
\label{D-3}
\end{equation}
where $D_{c2}$ again depends on $\tilde E_C$. For
$\tilde E_C \approx e^2/2d$
we have $D_{c2}^{-1}=2\pi\sqrt{2N_0}/\alpha$. Hence, the
dependence of $\tau_{\varphi 0}$ on $D$ for strongly disordered or
granular conductors (\ref{D-3}) is {\ it qualitatively} different
from that for sufficiently clean metals (\ref{D3}).

One can also roughly estimate the crossover between the regimes
$(a)$ and $(b)$ by requiring the values of $D=\gamma v_Fd/4$
(\ref{vFd}) and $D=g_c/4\pi N_0d$ to be of the same order. This
condition yields $(p_Fd)^2 \sim 2\pi g_c/\gamma$, and we arrive at
the estimate for $D$ at the crossover
\begin{equation}
D \approx \frac{0.6\hbar}{m} \sqrt{\frac{g_c}{\gamma}}.
\label{estD}
\end{equation}
Here we restored the Planck constant $\hbar$ set equal to unity
elsewhere in our paper.

In the next section we will use the above results and carry out a
detailed comparison between our theory and numerous available
experimental data for $\tau_{\varphi 0}$ in different types of
disordered conductors.

\section{Comparison with experiments}

Turning to experiments, it is important to emphasize again that
low temperature saturation of the electron decoherence time has
been repeatedly observed in numerous experiments and is presently
considered as firmly established and indisputably existing
phenomenon.  At the same time, the physical origin of this
phenomenon still remains under debate. The key observations and
remaining controversies are briefly summarized below.

\begin{enumerate}

\item{The authors \cite{Hackens} have analyzed the values of
$\tau_{\varphi 0}$ observed in their experiments with open quantum
dots as well in earlier experiments by different groups
\cite{Bird1,Clarke,Pivin,Huibers}. For all 14 samples reported in
\cite{Bird1,Clarke,Pivin,Huibers,Hackens} the values
$\tau_{\varphi 0}$ were found to rather closely follow a simple
dependence
\begin{equation}
\tau_{\varphi 0}\approx\tau_D.
\label{scdots}
\end{equation}
This {\it approximate} scaling was observed within the interval of
dwell times $\tau_D$ of about 3 decades, see Fig. 5 in Ref.
\cite{Hackens}. To the best of our knowledge, until now no
physical interpretation of this observation has been suggested.}

\item{In some of our earlier publications
\cite{GZ98,GZ99,Lammi,GZ02} we have demonstrated a good
quantitative agreement between our theoretical predictions
\cite{GZ98,GZ99} and experimental data for $\tau_{\varphi 0}$
obtained for numerous metallic wires and quasi-1d semiconductors.
As our theory of dephasing by electron-electron interactions
\cite{GZ98,GZ99} predicts a rather steep increase of
$\tau_{\varphi 0}$ with the system diffusion coefficient $D$, e.g.
for most metals as  $\tau_{\varphi 0} \propto D^3$, we can
conclude that for a large number of disordered conductors
$\tau_{\varphi 0}$ strongly {\it increases} with increasing $D$.}

\item{In a series of papers, see, e.g., Refs.
\cite{BL,Lin07,Lin01,Lin07b}, Lin and coworkers analyzed numerous
experimental data for $\tau_{\varphi 0}$ obtained by various
groups in disordered conductors with $D \lesssim 10$ cm$^2$/s and
observed systematic {\it decrease} of $\tau_{\varphi 0}$ with
increasing $D$. The data could be rather well fitted by the
dependence $\tau_{\varphi 0} \propto D^{-\alpha}$ with the power
$\alpha \gtrsim 1$. This trend is {\it opposite} to one observed
in less disordered conductors with $D \gtrsim 10$ cm$^2$/s and
remained unexplained until now.}

\item{The authors \cite{AAG} pointed out a disagreement between
our expressions for $\tau_{\varphi 0}$ \cite{GZ98,GZ99} and the
data obtained for a number of typically rather strongly disordered
2d and 3d structures. In some cases this disagreement was argued
to be as large as 4 to 5 orders of magnitude. In Ref. \cite{GZRep}
we countered this critique showing, on one hand, reasonable
agreement for some of the samples in question and arguing, on the
other hand, that our quasiclassical theory \cite{GZ98,GZ99} is
applicable merely to weakly disordered conductors. Hence, it
cannot be used in order to quantitatively describe strongly
disordered structures like, e.g., granular metals, metallic
glasses etc. quoted in Ref. \cite{AAG}. Formally eliminating the
controversy, this our argument, however, did not yet allow to
clarify the issue of low temperature saturation of $\tau_\varphi$
in strongly disordered conductors which remained unclear until
now.}

\item{Pierre {\it et al.} \cite{Saclay} argued that low
temperature saturation of $\tau_{\varphi}$ in sufficiently clean
samples can be caused by undetectably small number of magnetic
impurities. This idea, however, was not supported by the authors
\cite{MW} who demonstrated that at least in their experiments
(performed in high magnetic fields in order to fully polarize any
magnetic moments should they exist) the observed
$\tau_{\varphi}$-saturation cannot be due to magnetic impurities.
Later the issue was reanalyzed in Refs.
\cite{Gre,Gre-new,Birge,many} on the basis of recently developed
numerical renormalization group (NRG) theory of electron
scattering by Kondo impurities \cite{Zarand}. In samples with
implanted magnetic impurities a very good agreement between theory
\cite{Zarand} and experiments \cite{Gre-new,Birge,many} was
demonstrated both above and below the Kondo temperature $T_K$.
However, at $T \lesssim 0.1 T_K$ significant deviations from NRG
predictions was observed and $\tau_\varphi$ was found to saturate
\cite{Gre-new,Birge,many} in all samples both with and without
magnetic impurities. Interpretation of this saturation effect in
terms of both underscreened and overscreened models is problematic
\cite{Gre-new,many}}.

\end{enumerate}

Let us now analyze the above problems and controversies point by
point within the framework of our theory of electron-electron
interactions.

We start from the case of quantum dots
\cite{Bird1,Clarke,Pivin,Huibers,Hackens} observe that our results
for $\tau_{\varphi 0}$ (\ref{tauphi}), (\ref{t1}) scale
practically linearly with the dwell time $\tau_D$ which is
essentially the scaling (\ref{scdots}) suggested in Ref.
\cite{Hackens}. There is, however, one point which requires a
comment.

As it was argued in Sec. 2, one should expect no dephasing by
electron-electron interactions in single quantum dots at any $T$.
Naively one could regard this statement as yet one more
controversy with experiments
\cite{Bird1,Clarke,Pivin,Huibers,Hackens} where electron dephasing
in {\it single} dots was clearly observed. At this stage let us
recall that qualitative arguments in Sec. 2 as well as rigorous
analysis in Refs. \cite{GZ041,Brouwer} remain applicable to single
dots provided fluctuating voltages drop strictly across the two
barriers. In realistic quantum dots
\cite{Bird1,Clarke,Pivin,Huibers,Hackens} fluctuating voltages
most likely penetrate inside rather than drop only at the edges.
Within the framework of our model one can easily mimic this
situation by introducing additional scatterers inside the dot in
which case fluctuating voltages already do dephase (see Sec. 2).

For illustration, let us consider a strongly asymmetric
double dot system, i.e. we replace, say, the left barrier in Fig. 1
by a small quantum dot with the electron flight time
$\tau_{\rm fl}\ll \tau_D$. The left dot will then model
a barrier of a finite length in a single dot configuration of Fig. 1.
Applying Eqs. (\ref{GWLF1}), (\ref{WLF}) and (\ref{WLK}), we evaluate
the WL correction and again arrive at Eqs. (\ref{tT}-\ref{t0}), where,
however, $u=\tau_{\rm fl}/\tau_{RC}$. Extracting the decoherence time
$\tau_{\varphi 0}$, we obtain
\begin{equation}
\tau_{\varphi 0}\approx \frac{\tau_D}{(\tau_{\rm fl}/2\tau_{RC})^{2/g_Z}-1}
\label{tasym}
\end{equation}
for $\tau_{\rm fl}\gtrsim \tau_{RC}$ and vanishing decoherence rate in the
opposite limit $\tau_{\rm fl}\ll \tau_{RC}$. Provided the denominator
in Eq. (\ref{tasym}) is not very large, we arrive at  Eq. (\ref{scdots}).
We also note that $\tau_{\varphi 0}$ in a single dot (\ref{tasym}) is always
longer than that in a system of two dots (\ref{tauphi0}).

Alternatively, one can model a dot by a chain of several ($N$)
scatterers with $\tau_{\varphi }(T)$ defined in Eqs.
(\ref{t1}), (\ref{tauT}), (\ref{AAK}). Then at higher temperatures
we obtain $\tau_{\varphi } \propto T^{-\nu}$ with $\nu$ ranging
from 2/3 to 1, as observed in a number of dots
\cite{Bird1,Clarke,Pivin,Huibers,Hackens}. Substituting $g\tau_D$
by $g_{\rm tot}\tau_D^{\rm tot}/N$ (where $g_{\rm tot}$ are
$\tau_D^{\rm tot}$ are respectively the dimensionless conductance
and the dwell time of a composite dot), bearing in mind that
 $g_{\rm tot} \approx 1\div 12$ \cite{Hackens} and
assuming $N$ to be not very large, at low $T$ from (\ref{t1}) one
finds $\tau_{\varphi 0} \sim \tau_D^{\rm tot}$ in agreement with
experimental results presented in Fig. 5 of Ref. \cite{Hackens}.
More complicated configurations of scatterers can also
be considered with essentially the same results. We conclude that
our present theory is in a good agreement with experimental
findings \cite{Bird1,Clarke,Pivin,Huibers,Hackens}.

Let us now turn to experiments with spatially extended conductors.
In Fig. 6 we have collected experimental data for $\tau_{\varphi
0}$ obtained in over 120 metallic samples with diffusion
coefficients varying by $\sim 4$ decades, from $D \approx 0.3$
cm$^2$/s to $D \approx 350$ cm$^2$/s. The data were taken from
about 30 different publications listed in figure caption. We see
that the the measured values of $\tau_{\varphi 0}$ strongly depend
on $D$. Furthermore, this dependence turns out to be
non-monotonous: For relatively weakly disordered structures with
$D \gtrsim 10$ cm$^2$/s $\tau_{\varphi 0}$ clearly increases with
increasing $D$, while for strongly disordered conductors with $D
\lesssim 10$ cm$^2$/s the opposite trend takes place. In addition
to the data points in Fig. 6 we indicate the dependencies
$\tau_{\varphi 0} (D)$ (\ref{D3}) and (\ref{D-3}) for two models
$(a)$ and $(b)$ discussed in Sec. 5.2.

\begin{figure}
\includegraphics[width=7.5cm]{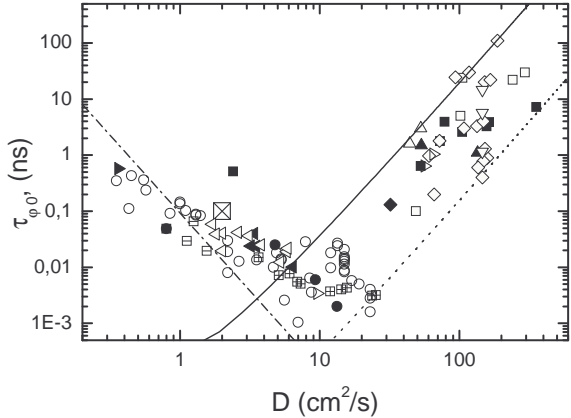}
\caption{The low temperature dephasing times observed in various
experiments: Au \cite{Ba},\cite{Gre} ($\square$);
Au \cite{MW} ($\triangle$); Au,Ag \cite{Saclay} ($\lozenge$);
Au Pd \cite{Lin07} ($\circ$);
\cite{Bird} ($\bullet$); Au \cite{Birge} ($\triangledown$);
\cite{Sah} ($\boxtimes$, ten points within the box);
\cite{Lin01} Au$_2$Al ($\blacktriangleleft$), Sb ($\triangleleft$),
Sc$_{85}$Ag$_{15}$ ($\blacktriangleright$), V$_3$Al ($\boxminus$);
\cite{Nat} ($\triangleright$); \cite{Lin07b} CuGeAu ($\boxplus$);
AuPd-3 and Al \cite{Altomare} ($\blacklozenge$) Au \cite{Enh}($\blacktriangle$);.}
\end{figure}

We observe that for  $D \gtrsim 10$ cm$^2$/s the data points
clearly follow the scaling (\ref{D3}). Practically all data points
remain within the strip between the two lines corresponding to Eq.
(\ref{D3}) with $\gamma = 1$ (dashed line) and $\gamma = 0.2$
(solid line). On the other hand, for more disordered conductors
with $D \lesssim 10$ cm$^2$/s the data are consistent with the
scaling (\ref{D-3}) obtained within the model $(b)$. We would like
to emphasize that theoretical curves (\ref{D3}) and (\ref{D-3})
are presented in Fig. 6 without any additional fit parameters
except for a geometry factor $\gamma$ for the first dependence and
the value $g_c \approx 150$ for the second one. This value of
$g_c$ was estimated from the crossover condition (\ref{estD}) with
$D \sim 10$ cm$^2$/s and $\gamma \sim 1$.

Note that quite a few data points with $D \gtrsim 10$ cm$^2$/sek
correspond to the samples contaminated by magnetic impurities.
Remarkably, these data points also demonstrate -- though with
somewhat larger scatter -- {\it systematic} increase of
$\tau_{\varphi 0}$ with increasing $D$. At the same time, for
similar values of $D$ the samples with higher concentration of
magnetic impurities have systematically lower $\tau_{\varphi 0}$
than samples with few or no magnetic impurities. These
observations indicate that for samples with relatively high
concentration of magnetic impurities both mechanisms of
electron-electron interactions and spin-flip scattering provide
substantial contributions to $\tau_\varphi$, being responsible
respectively for the scaling $\tau_{\varphi 0} \propto D^3$ and
for additional non-universal shift of the data points downwards.

In order to carry out more accurate comparison with our theory of
electron-electron interactions let us now leave out the data
points taken from the samples with high concentration of magnetic
impurities. In Fig. 7 we selected the data for 37 different
metallic wires with no or few magnetic impurities and with
diffusion coefficients in the range 9 cm$^2$/s $<D< 300 $cm$^2$/s
from Refs.
\cite{Moh,MJW,Nat,Saclay,MW,Ba,Bird,Nat05,Gre,Gre-new,Birge,many,Mohunp,Altomare}.
Two minor adjustments of some data points are in order. Firstly,
in order to eliminate the uncertainty related to different
definitions of $\tau_\varphi$ used by different groups
\cite{FN01}, we have adjusted 6 data points \cite{Nat} and 2 data
points \cite{Bird} according to the definition of $\tau_\varphi$
used in other works \cite{Moh,MJW,Saclay,Nat05,Gre,Gre-new,Birge}.
Secondly, 5 data points corresponding to samples with high
diffusion coefficients \cite{Saclay}  have been adjusted in order
to eliminate the temperature dependent contribution to
$\tau_\varphi$ which remains substantial for these samples down to
the lowest $T \approx 40$ mK \cite{FN02}. Both these adjustments
can in no way influence our conclusions (in fact, non-adjusted
data points also remain in-between solid and dashed lines in Fig.
7).

\begin{figure}
\includegraphics[width=7.5cm]{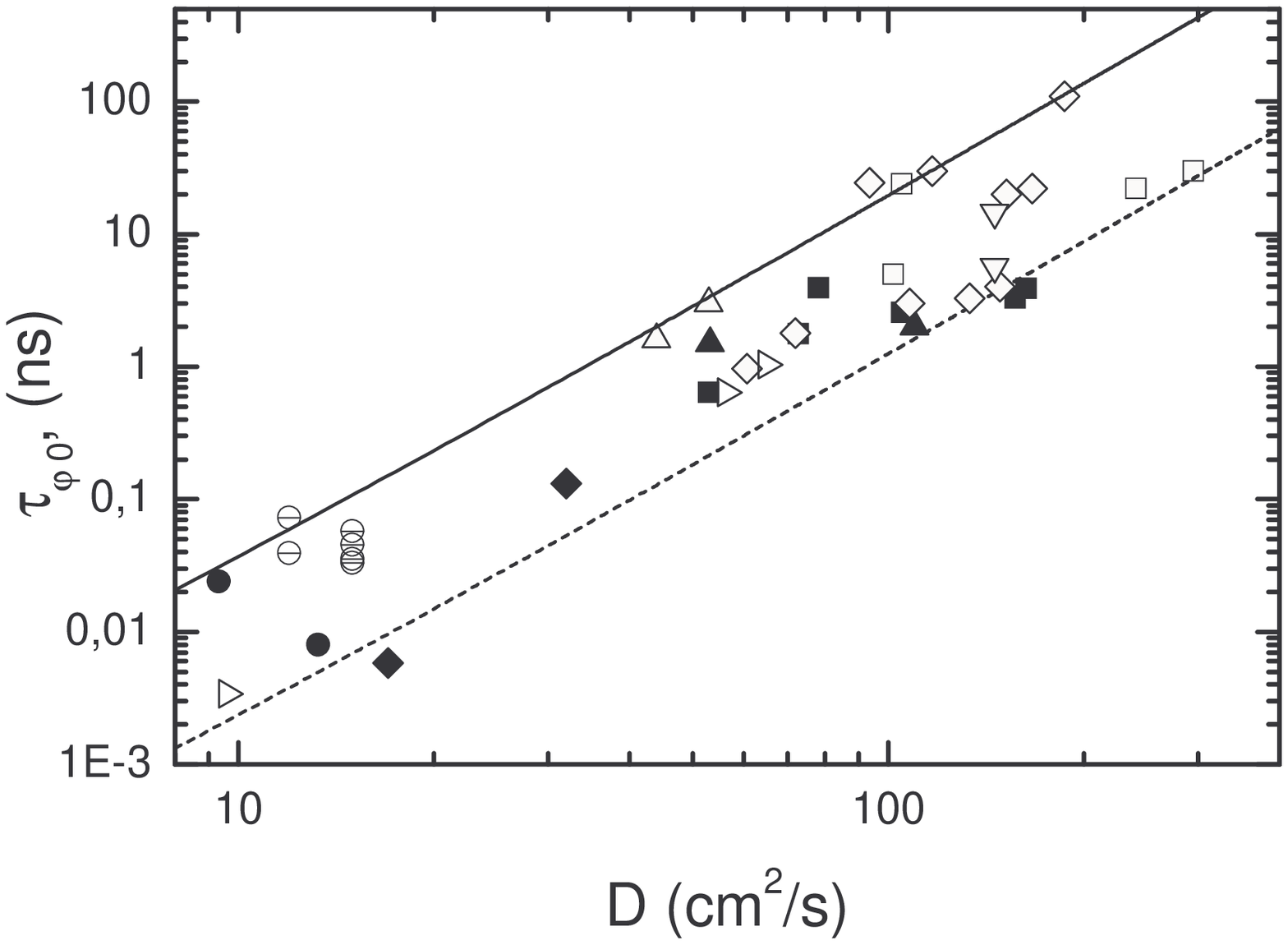}
\caption{The same as in Fig. 6 for 9 cm$^2$/s $<D< 300 $cm$^2$/s.
Only the data for the samples with low concentration or with no
magnetic impurities have been selected: Au-1, Au-3, Au-4, Au-6
\cite{Moh}, Au-7 \cite{MJW}, Au-9 \cite{Mohunp} and CF-1, CF-2
\cite{MW} ($\blacksquare$); A to F (AuPd) \cite{Nat}
$\circleddash$); Ag(6N)a,b,c,d, Au(6N), Ag(5N)a,b, and Cu(6N)a,b
\cite{Saclay} ($\lozenge$); Au \cite{Ba}, Au1\cite{Gre}, Ag1
\cite{many} and Ag2 \cite{Gre-new} ($\square$); 2 samples from
\cite{Bird}, still to be included, 2 Au \cite{Birge}
($\triangledown$); D (AuPd), F, H (Ag) \cite{Nat05}
($\triangleright$); AuPd-3 and Al \cite{Altomare}($\blacklozenge$)
Au \cite{Enh} ($\blacktriangle$).}
\end{figure}

The data in Fig. 7 again clearly support the scaling of
$\tau_{\varphi 0}$ with $D$ (\ref{D3}) due to electron-electron
interactions. Practically all data points are now located
in-between the solid and dashed lines which indicate the
dependence (\ref{D3}) with respectively $\gamma =0.2$ and $\gamma
=0.5$. We cannot exclude that the remaining scatter among the data
points with similar values of $D$ -- to a certain extent -- may be
due to relatively small amount of magnetic impurities possibly
residing in some samples. Also minor differences in metal
parameters (e.g, Fermi velocities) may contribute to this effect.
It appears, however, that sample-to-sample fluctuations of the
geometry parameter $\gamma$ play the most important role. This
parameter is defined as a ratio between the square root of the
inter-dot (inter-grain) contact area $\mathcal{A}^{1/2}$ and the
average dot (grain) size $\mathcal{V}^{1/3}$, i.e. $\gamma$
accounts for {\it local} properties of the sample which can be
highly non-universal. Electrons in metallic wires get scattered
both in the bulk and on the surface. Hence, e.g., the surface
quality and/or details of the wire geometry -- along with other
factors -- may significantly impact the value of $\gamma$. Since
$\tau_{\varphi 0}$ (\ref{D3}) depends quite strongly on $\gamma$,
it is by no means surprising that different wires with similar
values of $D$ may have decoherence times differing by few times.

As an illustration of this point let us consider 4 quasi-1d AuPd
samples C, D, E, F \cite{Nat} with nominally identical values of
$D=15$ cm$^2$/s. Although these samples were fabricated in the
same way and measured in one experiment \cite{Nat}, their
dephasing times were found to differ by up to $\sim 2$ times. This
difference can hardly be ascribed to magnetic impurities since the
measured level of dephasing would require unrealistically high
concentration of such impurities (in the range of 100 ppm) and, in
addition, rather exotic values of the Kondo temperature $T_K$. At
the same time, sample-to-sample fluctuations of the parameter
$\gamma$ of only $\lesssim 25$ per cent (which is easy to assume
given, e.g., somewhat different geometry of the samples) would
fully account for the above difference of dephasing times. We
conclude that our theoretical expression (\ref{D3}) is in a good
quantitative agreement with the available experimental data for
quasi-1d metallic wires with diffusion coefficients in the range 9
cm$^2$/s $<D< 300$ cm$^2$/s.

Now let us return to Fig. 6 and consider the data for strongly
disordered conductors with  $D< 10 $ cm$^2$/s. As we already
pointed out, the agreement between the data and the dependence
(\ref{D-3}) predicted within our simple model $(b)$ is reasonable,
in particular for samples with $D< 3 $cm$^2$/s. At higher
diffusion coefficients most of the data points indicate a weaker
dependence of $\tau_{\varphi 0}$ on $D$ which appears natural in
the vicinity of the crossover to the dependence (\ref{D3}). The
best fit for the whole range 0.3 cm$^2$/s $<D< 10$ cm$^2$/s is
achieved with the function $\tau_{\varphi 0} \propto D^{-\alpha}$
with the power $\alpha \approx 1.5 \div 2$. Further modifications
of our -- clearly oversimplified -- model $(b)$ can help to
achieve even better agreement between theory and experiment.

Although such modifications can certainly be worked out, it is not
our aim to do it here. More importantly, our analysis of Sec. 5.2
allows to qualitatively understand and explain seemingly
contradicting dependencies of $\tau_{\varphi 0}$ on $D$ observed
in weakly and strongly disordered conductors. While the trend
``less disorder -- less decoherence'' (\ref{D3}) for sufficiently
clean conductors is quite obvious, the opposite trend  ``more
disorder -- less decoherence'' in strongly disordered structures
requires a comment. Effectively the latter dependence implies that
with increasing disorder electrons spend more time in the areas
with fluctuating in time but spatially uniform potentials which do
not dephase, as we also discussed in Sec. 2. In other words, in
this case an effective dwell time $\tau_D$ in Eq. (\ref{t1})
becomes longer with increasing disorder and, hence, the electron
decoherence time $\tau_{\varphi 0}$ does so too.

Is the physical picture of $D$ {\it decreasing} with increasing
dot (or grain) size (employed within the model $(b)$) realistic?
Although for cleaner conductors the tendency is usually just the
opposite, for strongly disordered structures increasing
resistivity with increasing grain size has been observed in a various
experiments \cite{ind,cor,cor2}.
In addition, since local conductance fluctuations increase with
increasing disorder, several grains can form a cluster with
internal inter-grain conductances strongly exceeding those at its
edges. In this case fluctuating potentials remain almost uniform
inside the whole cluster which will then play a role of an
effective (bigger) grain/dot. Accordingly, the average volume of
such ``composite dots'' $\mathcal{V} \propto 1/\delta$ may grow
with increasing disorder, electrons will spend more time in such
bigger dots and, hence, the electron decoherence time (\ref{t1})
will increase.

The above comparison with experiments confirms that our previous
quasiclassical results for $\tau_{\varphi 0}$
\cite{GZ98,GZ99,GZ00} are applicable to relatively weakly
disordered structures with $D\gtrsim 10$ cm$^2$/s, while for
conductors with stronger disorder different expressions for
$\tau_{\varphi 0}$ (e.g., Eq. (\ref{D-3})) should be used. For
instance, the claimed in Ref. \cite{AAG} ``disagreement'' between
our theory and some experiments by 4 to 5 orders of magnitude is
solely due to misuse of our results \cite{GZ98,GZ99} far beyond
their applicability range. A glance at Fig. 6 is sufficient to
realize that any attempt to apply Eq. (\ref{D3}) to structures
with, e.g., $D \sim 0.3\div 1$ cm$^2$/s can easily lead to a
``disagreement'' with the data, say, by 6 orders of magnitude or
so. This observation, however, does not imply any real
disagreement, rather it indicates that more general results for
$\tau_{\varphi 0}$, e.g., Eq. (\ref{t1}), should be used. The same
argument invalidates the comparison between our quasiclassical
results \cite{GZ98,GZ99} and the data for the sample Au-5
\cite{Moh}, see Fig. 8 of Ref. \cite{Saclay}, or, to a somewhat
lesser extent, for the sample C of Ref. \cite{Nat05}, see Table II
of that paper \cite{FN03}.

Finally, our analysis allows to rule out scattering on magnetic
impurities as a cause of low temperature saturation of
$\tau_\varphi$. This mechanism can explain neither strong and
non-trivial dependence of the electron decoherence time on $D$
(see our Figs. 6 and 7 as well as, e.g., Fig. 4 in \cite{Nog},
Fig. 1 in \cite{Lin07} and Fig. 5 in \cite{Hackens}) nor even the
level of dephasing observed in numerous experiments. E.g., in
order to be able to attribute dephasing times as short as
$\tau_{\varphi 0} \lesssim 10^{-12}$ s to magnetic impurities one
needs to assume huge concentration of such impurities ranging from
few {\it hundreds} to few {\it thousands} ppm which appears highly
unrealistic, in particular for systems like carbon nanotubes,
2DEGs or quantum dots. Similar arguments were independently
emphasized by Lin and coworkers \cite{Lin07,Lin07b}. Even in
metallic wires with high values of $D$ and long dephasing times,
at $T \lesssim 0.1 T_K$ one observes clear saturation of
$\tau_\varphi$ \cite{Gre,Gre-new,Birge} which is in a {\it
quantitative} agreement with our theory of electron-electron
interactions (see Fig. 7) and is very hard to explain otherwise
\cite{many}.

Thus, although electron dephasing due to scattering on magnetic
impurities is by itself an interesting issue, its role in low
temperature saturation of $\tau_\varphi$ in disordered conductors
is sometimes strongly overemphasized. Since the latter phenomenon
has been repeatedly observed in {\it all} types of disordered
conductors, the physics behind it should most likely be universal
and fundamental. We believe -- and have demonstrated here -- that
it is indeed the case: Zero temperature electron decoherence in
all types of conductors discussed above is caused by
electron-electron interactions.

\section{Conclusions}
In this paper we have employed a model of an array of quantum
dots/scatterers (Fig. 3) which embraces various types of
disordered conductors and allows to study electron transport in
the presence of interactions within a very general theoretical
framework. We have non-perturbatively analyzed the impact of
electron-electron interactions on weak localization for such
structures with the emphasis put on the interaction-induced
decoherence of electrons at low temperatures. We have formulated a
fully self-contained theory free of any divergencies and cutoffs
which allows to conveniently handle disorder averaging and treat
electron scattering without residing to quasiclassics. In the case
of two quantum dots (or three scatterers) it was possible to find
an exact solution of the problem (Sec. 4) and to evaluate the WL
correction to the system conductance practically without
approximations.

With the aid of our approach we have formulated a unified
description of electron dephasing by Coulomb interaction in
different structures including (i) weakly disordered conductors
(e.g., metallic wires with $D \gtrsim 10$ cm$^2$/s), (ii) strongly
disordered conductors ($D \lesssim 10$ cm$^2$/s) and (iii)
metallic quantum dots. We have demonstrated that in all these
cases at $T \to 0$ the electron decoherence time is determined by
the same simple formula $\tau_{\varphi 0}\sim g\tau_D/\ln
(E_C/\delta)$. In the case (i) this formula yields $\tau_{\varphi
0}\propto D^3/\ln D$ and matches with our previous quasiclassical
results \cite{GZ98,GZ99,GZ00} while in the cases (ii) and (iii) it
illustrates new physics which was not yet explored before. In
particular, this formula emphasizes the dependence of
$\tau_{\varphi 0}$ on the electron dwell time $\tau_D$ in single
quantum dots \cite{Hackens} and helps to understand the (at the
first sight counterintuitive) trend ``more disorder -- less
decoherence'' observed in strongly disordered conductors
\cite{BL,Lin07,Lin01,Lin07b}.

We have carried out a detailed comparison of our theoretical
predictions with the results of numerous experiments for the whole
scope of structures (i), (ii) and (iii) (Sec. 6). In all cases we
found a good agreement between theory and experiment which further
supports our main conclusion that low temperature saturation of
$\tau_\varphi$ is universally caused by electron-electron
interactions.

\section*{Acknowledgments}

We are grateful to C. Bauerle, J. Bird, P. Hakonen, P. Mohanty, D.
Natelson, M. Paalanen, J.-J. Lin, L. Saminadayar and R.A. Webb for
useful discussions on various experimental aspects of electron
decoherence and/or for providing us with their experimental data.

This work is part of the European Community's Framework Programme
NMP4-CT-2003-505457 ULTRA-1D "Experimental and theoretical investigation of
electron transport in ultra-narrow one-dimensional nanostructures".


\begin{thebibliography}{10}
\bibitem{FN01} Different groups use different functions in order
to fit their magnetoresistance data. As a result the values of
$\tau_\varphi$ extracted from such fits may differ by a numerical
prefactor. Here we stick to the standard definition of
$\tau_\varphi$ based on, e.g., Eq. (1) of Ref. \cite{Saclay}. With
this choice, the values of $\tau_\varphi$ obtained from fits with
the Airy function \cite{Nat,Bird} should be ajusted (moved
upwards) by a numerical factor $\sim 4$. For further discussion of
this point we refer the reader to Appendix B in \cite{Saclay}. We
also note that the adjustment of the data points \cite{Nat,Bird}
was carried out only in Fig. 7 but not in Fig. 6.
\bibitem{FN02} According to our result (\ref{D3}) $\tau_{\varphi
0}$ can be very long at large values of $D$. Hence, saturation of
$\tau_\varphi$ in such samples can only be observed at much lower
temperatures as compared to samples with smaller values of $D$.
E.g., the samples Ag6Na,b,c,d and Au6N reported in Ref.
\cite{Saclay} down to temperature $T \sim 40$ mK were still not in
the true saturation regime. Later the samples with similar values
of $D$ were cooled down to lower temperatures \cite{Gre,Gre-new}
and saturation of $\tau_\varphi$ was indeed observed at $T
\lesssim 30$ mK, see Fig. 6 of Ref. \cite{many}. Thus, in order 
to perform a fair comparison, the values
$\tau_{\varphi 0}$ for the samples  Ag6Na,b,c,d and Au6N
\cite{Saclay} should be determined by fitting the measured
temperature dependencies for $\tau_\varphi$ to our theory. This
procedure was described in detais in Ref. \cite{GZ02}.
\bibitem{FN03} The data for the samples D, G and H \cite{Nat05}
agree well with our theory \cite{GZ98,GZ99}, as is also seen in
Fig. 7. Note that the (saturated) experimental values of
$L_\varphi$ for the samples G and H are respectively 1.9 $\mu$m
and 2.6 $\mu$m \cite{Nat07}. Their ratio $\approx 0.73$ matches
perfectly with our theoretical prediction $\approx 0.75$, cf.
Table II in \cite{Nat05}.
\end{thebibliography}
\end{document}